\documentclass[lettersize,journal]{IEEEtran}
\normalsize
\ifCLASSINFOpdf
   \usepackage[pdftex]{graphicx}
   \graphicspath{{../pdf/}{../jpeg/}}
   \DeclareGraphicsExtensions{.pdf,.jpeg,.png}
\else
   \usepackage[dvips]{graphicx}
   \graphicspath{{../eps/}}
   \DeclareGraphicsExtensions{.eps}
\fi
\usepackage{cite}
\usepackage[colorlinks,linkcolor=blue,anchorcolor=blue,citecolor=blue]{hyperref}
\usepackage{bm}
\usepackage{subfigure}
\usepackage{dsfont}		
\usepackage{multirow}
\usepackage{makecell}
\usepackage{color}
\usepackage{arydshln}  
\usepackage{amsmath}   
\usepackage{amssymb}   
\usepackage{amsfonts}
\usepackage{mathrsfs}  
\usepackage[linesnumbered,ruled]{algorithm2e}
\newtheorem{definition}{Definition}
\newtheorem{example}{Example}
\newtheorem{remark}{Remark}
\newtheorem{theorem}{Theorem}
\newtheorem{lemma}{Lemma}

\newtheorem{proposition}{Proposition}

\makeatletter
\newcommand{\biggg}{\bBigg@{3}}

\makeatother
\SetCommentSty{myCommentStyle}	

\begin{document}
	
\title{Protograph-Based Batched Network Codes}

\author{
	Mingyang~Zhu,~\IEEEmembership{Member,~IEEE,} Ming~Jiang,~\IEEEmembership{Member,~IEEE,} and~Chunming~Zhao,~\IEEEmembership{Member,~IEEE}
	\thanks{M. Zhu was with the National Mobile Communications Research Laboratory, Southeast University, Nanjing 210096, China. He is now with the Institute of Network Coding, The Chinese University of Hong Kong, Hong Kong SAR, China (e-mail: mingyangzhu@cuhk.edu.hk).}
	\thanks{M. Jiang and C. Zhao are with the National Mobile Communications Research Laboratory, Southeast University, Nanjing 210096, China, and also with the Purple Mountain Laboratories, Nanjing 211111, China (e-mail: jiang\_ming@seu.edu.cn; cmzhao@seu.edu.cn).}
}
\maketitle

\begin{abstract}
	Batched network codes (BNCs) are a low-complexity solution for communication through networks with packet loss. Although their belief propagation (BP) performance is proved to approach capacity in the asymptotic regime, there is no evidence indicating that their BP performance is equally good in the finite-length regime. In this paper, we propose a protograph-based construction for BNCs, referred to as protograph-based BNCs (P-BNCs), which significantly differs from existing BNCs in three aspects: 1) The vast majority of existing construction methods mainly focus on the degree distribution of check nodes (CNs), whereas P-BNCs not only specify the degree distributions of CNs and variable nodes (VNs) but also partially constrain the connectivity between CNs and VNs. 2) Traditional BNCs use a fixed degree distribution to generate all batches, making their performance highly sensitive to channel conditions, but P-BNCs achieve good performance under varying channel conditions due to their rate-compatible structures. 3) The construction of P-BNCs takes into account joint BP decoding with a sparse precode, whereas traditional constructions typically do not consider a precode, or assume the presence of a precode that can recover a certain fraction of erasures. Due to these three improvements, P-BNCs not only have higher achievable rates under varying channel conditions, but more importantly, their BP performance is significantly improved at practical lengths.
\end{abstract}

\IEEEpeerreviewmaketitle

\section{Introduction}
Batched network codes (BNCs) (a.k.a. chunked, segmented, or generation-based network codes) are a class of low-complexity network codes for reliable communication through networks~\cite{maymounkov2006methods,5191397,overlapChunk,5695118,GammaCodes2012,Tang2012EOC,BATS,Tang2018Lchunked}. In these coding schemes, a \textit{batch} is a small set of packets within which network coding is applied to the packets belonging to the same batch. Such a batched structure reduces the computational complexity of encoding and decoding, the storage complexity of buffered packets, and the overhead of coefficient vectors~\cite{BATS}. To achieve higher achievable rate while maintaining low decoding complexity, \textit{sparse linear constraints} are typically imposed to batches, allowing message passing among batches through belief propagation (BP). Overlapping is a straightforward way to establish connections between batches, e.g.,~\cite{5191397,overlapChunk,5695118}. Applying an outer code or precode is another general method to impose constraints on batches, e.g.,~\cite{GammaCodes2012,BATS,Tang2018Lchunked}.

Batched sparse (BATS) codes are a notable class of BNCs which have garnered significant attention in recent years due to their low complexity and high achievable rates~\cite{BATS,Yang2016tree,FL_analysis_BATS,Xu2017QUBATS,8013842,8629008,9594262,9664430,Zhu2023LDPCBATS}. Many studies have demonstrated that BATS codes under BP decoding can approach capacity \textit{in the asymptotic regime}~\cite{BATS,Yang2016tree}. However, BP performance of BATS codes \textit{in the finite-length regime}\footnote{In this paper, ``finite length'' does not imply an extremely small length, but rather is equivalent to ``practical length''. For example, we might be interested in lengths ranging from a few hundred to a few thousand.} is not very satisfactory, especially when the codes are designed based on asymptotic analysis~\cite{FL_analysis_BATS,8629008,8013842,9664430,qing2024dependence}.\footnote{It should not be misconstrued that the finite-length performance of other BNCs, besides BATS codes, is good. In fact, the finite-length performance of other BNCs is also unsatisfactory, e.g., see~\cite{5695118,overlapChunk,GammaCodes2012}. Additionally, many studies on other BNCs discuss only the asymptotic performance without providing numerical results on finite-length performance, e.g.,~\cite{Tang2012EOC,Tang2018Lchunked}.} In~\cite{FL_analysis_BATS}, the authors found that the BATS codes optimized by finite-length analysis significantly outperform those optimized by asymptotic analysis when the number of input packets is $256$ (the performance curve of asymptotically optimal BATS codes of this length is very flat, somewhat similar to an error floor). In~\cite{8013842} and~\cite{8629008}, two variants of BATS codes were proposed which are called expanding-window BATS (EW-BATS) codes and sliding-window BATS (SW-BATS) codes, aiming to protect input packets unequally or reduce the transmission latency, but the BP performance of these two variants are still far away from the theoretical limit. Later, EW-BATS codes have been improved in~\cite{9664430}, called adaptive EW-BATS (AEW-BATS) codes. However, the improvement on error probability made by AEW-BATS codes is mainly observed when the transmission overhead is large.

Recently, some works have focused on improving the finite-length performance of BATS codes based on low-density parity-check~\cite{Gallager} \textit{precoding}. In~\cite{Zhu2023LDPCBATS}, a heuristic optimization problem was formulated according to the recursive finite-length analysis of BATS codes~\cite{FL_analysis_BATS} and the decoding threshold of LDPC codes~\cite{910577,modern_coding_theory}. In~\cite{zhu2024TIT}, rigorous finite-length analysis of LDPC-precoded BATS codes was made to optimize the code construction, thus significantly reducing the gap between the actual performance and the theoretical limit for short BATS codes. There may be two main reasons why LDPC precoding can enhance the performance of BATS codes (the explanations are based on the Tanner graph of the code, which will be introduced in Sec.~\ref{subsection:BATS}): 1) The LDPC precode imposes structured constraints to the variable nodes (VNs), i.e., packets, preventing the poor connectivity of some VNs that could result from the random property of the BATS code itself (in the worst case, not being connected to any batch). 2) The constraints of the LDPC precode result in near random interleaving among the VNs, which mitigate the insufficient interleaving caused by the ``batched interleaver'' of the BATS code (coded packets in the same batch share the same connections).

However, designing BATS codes based on finite-length analysis is computationally intensive and is generally only applicable to scenarios with a few hundred input packets. As mentioned before, BATS codes with thousands of input packets still do not achieve satisfactory BP performance. Therefore, we want to improve the existing asymptotic analysis-based design methods for BATS codes to achieve excellent BP performance in the finite-length regime (because the asymptotic analysis is, in general, much simpler than the finite-length analysis). In fact, the mainstream design methods for LDPC codes align with our expectation. Inspired by the protograph-based design commonly used in the LDPC literature~\cite{JPL,1523619,5174517,6266764,7045568,7112076,10639052} and the aforementioned idea of LDPC precoding, this paper proposes protograph-based BNCs (P-BNCs). In the protograph of BNCs, packets and batches correspond to VNs and check nodes (CNs), respectively. The construction process of P-BNCs is similar to that of protograph-based LDPC codes, which includes two steps: Optimizing the protograph and then lifting the protograph. To this end, several key issues need to be addressed: 1) How to design the protograph of P-BNCs and lift it to a code? 2) How to develop asymptotic analysis based on the protograph and define the decoding threshold? 3) How about the performance of such constructed P-BNCs in the finite-length regime?

\begin{figure}[!t]
	\centering
	\includegraphics[width=3in]{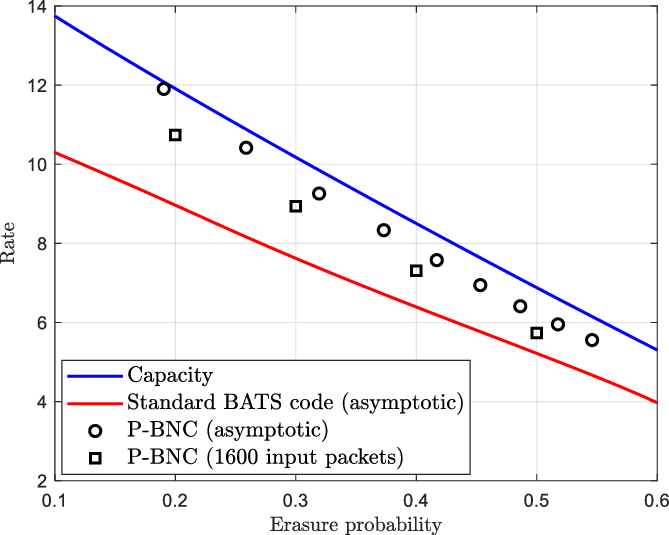}
	\caption{Achievable rates of the standard BATS code and the P-BNC over a line network with two channels. The two erasure probabilities of two channels are supposed to be the same. The batch size is $16$ (i.e., each batch has 16 coded packets). The standard BATS code is optimized for the erasure probability varying from $0.1$ to $0.6$ using \cite[(P3)]{BATS}. The rate is defined as $A/N$, where $A$ is the number of input packets and $N$ is the number of batches. To evaluate the performance at a practical length, we set $A = 1600$, and $N$ is chosen based on simulation at which the frame error rate reaches $0.1$.}
	\label{fig:achievable_rate}
\end{figure}

This paper addresses the aforementioned issues. The main results are summarized as follows.
\begin{enumerate}
	\item A BNC protograph structure combined with LDPC precoding is proposed, which has a rate-compatible structure that can adapt to channels under different conditions. Additionally, to construct codes based on a small protograph,\footnote{Consider a BNC with 250 packets and 10 batches. The protograph size must be chosen from $1 \times 25,2 \times 50,\ldots,10 \times 250$ (the number of CNs $\times$ the number of VNs). Empirically, it is necessary to ensure a certain number of CNs to achieve a better optimized protograph, which correspondingly leads to a quite large-size protograph.} a scheme for deleting specific batches from the lifted graph, referred to as \textit{puncturing}, is proposed. Furthermore, a framework for optimization of the protograph is presented. 
	\item Protograph-based asymptotic analysis for P-BNCs is developed, which is a non-trivial extension of the tree-based analysis in~\cite{Yang2016tree} to protographs. Due to the presence of multiple channels in the network, the decoding threshold can not be defined as some parameter of a single channel. We define the decoding threshold for a P-BNC over a multicast network based on \textit{network capacity}, where the erasure probabilities of all channels and the decodabilities of all destination nodes are considered. The decoding threshold is a new concept for BNCs and plays the key role in optimization of the protograph.
	\item The numerical results in this paper indicate that the optimized P-BNCs can achieve excellent performance under BP decoding. In particular, at the frame error rate (i.e., the probability that the decoder cannot recover \textit{all} input packets) of $0.1$, the gap to the theoretical limit can be reduced to about $5 \sim 20$ batches, equivalently, about $10\% \sim 20\%$ overhead. A representative result is shown in Fig.~\ref{fig:achievable_rate}, and more results are provided in Sec.~\ref{section:optimization}.
\end{enumerate}

The rest of this paper is organized as follows. Sec.~\ref{sec:preliminary} is a preliminary. Sec.~\ref{section:BNC} introduces BNCs and Sec.~\ref{section:PBNC} presents the protograph structure for BNCs. Sec.~\ref{sec:asy} develops asymptotic analysis for P-BNCs. Sec.~\ref{section:optimization} presents a framework for protograph optimization and provides numerical results. Sec.~\ref{sec:conclusion} summarizes this paper and discusses several open questions.

\section{Preliminaries}\label{sec:preliminary}

\subsection{Notation}
Let $\mathbb{F}$ denote a finite field containing $q$ elements. $[{\bf A},{\bf B}]$ represents the concatenation of two matrices by stacking them horizontally, i.e., $[{\bf A} \mid {\bf B}]$. $[{\bf A};{\bf B}]$ represents the juxtaposition of two matrices by stacking them vertically, i.e., \scalebox{0.7}{$\begin{bmatrix}{\bf A}\\{\bf B}\end{bmatrix}$}. The rank of $\bf A$ is denoted by ${\rm rk}({\bf A})$. ${\bf a}[i:j]$ denotes the subvector of ${\bf a}$ composed of the $i$-th to $j$-th elements. ${\bf A}[i,j]$ denotes the element in matrix $\bf A$ at the $i$-th row and $j$-th column.

\subsection{Protograph Codes}\label{subsection:protograph_codes}
\begin{figure*}[!t]
	\centering
	\includegraphics[width=\linewidth]{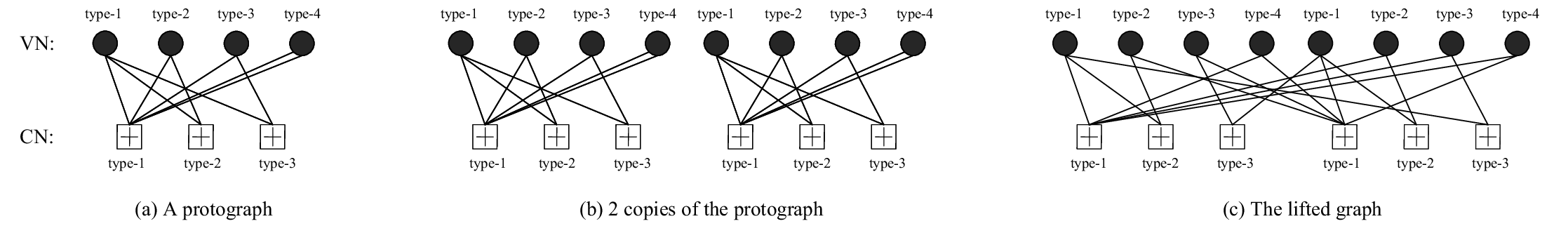}
	\caption{An example of lifting a protograph with $Z = 2$. The figure is referenced from~\cite{7112076}.}
	\label{fig:protograph_lift}
\end{figure*}
A protograph, which was first introduced in~\cite{JPL}, is a Tanner graph with a relatively small number of variable nodes (VNs) and check nodes (CNs)~\cite{1523619,5174517,6266764,7045568,7112076}. For a protograph having $n_v$ VNs and $n_c$ CNs, each node and each edge in the protograph are given a \textit{type}, say the type-$i$ CN, type-$j$ VN, and type-$(i,j)$ edge, where $i \in \{1,2,\ldots,n_c\}$ and $j \in \{1,2,\ldots,n_v\}$ (one can treat the type as a way of labeling the nodes and edges). Edges of the same type are allowed in the protograph (i.e., there may exist more than one edges between a CN and a VN, called \textit{parallel edges}). To obtain a lifted graph that defines a code, a \textit{lifting} (i.e., ``copy-and-permute'') operation can be applied to the protograph to expand the protograph by a factor of $Z$, where $Z$ is call the \textit{lifting factor}. First, the protograph is \textit{copied} $Z$ times, resulting in $Z$ individual protographs. Then, \textit{edge permutation} takes place among all edges in $Z$ copies of the protograph that have the same type. By edge permutation, the copies of the protograph become coupled, and all parallel edges are ``eliminated'' (be distributed into different copies of the protograph). After that, a lifted graph is obtained, where any type-$(i,j)$ edge still connects a type-$i$ CN and a type-$j$ VN. A protograph code is defined by the lifted graph, which has the same \textit{design rate} and the same \textit{degree distribution} of VNs and CNs as the protograph. Different lifting strategies may lead to different performance of the code. For $Z$ copies of a protograph, if all edges of each type are totally randomly permuted, this lifting operation is called a \textit{random lifting}.

A protograph can be represented by a non-negative integer matrix ${\bf B} = [b_{i,j}]_{1 \le i \le n_c, 1 \le j \le n_v}$, called the \textit{protomatrix}. The $i$-th row and the $j$-th column of the protomatrix ${\bf B}$ correspond to the type-$i$ CN and the type-$j$ VN in the protograph, respectively, and $b_{i,j}$ is the number of edges between the corresponding two nodes. Since there is no essential difference between protograph and protomatrix, we use them interchangeably. The following example illustrates the protograph, the protomatrix, and the lifting operation to obtain a code.

\begin{example}
	Fig.~\ref{fig:protograph_lift}(a) is a protograph consists of 4 VNs, 3 CNs, and 9 edges, including parallel edges between the type-1 CN and type-4 VN, and the corresponding protomatrix is given by
	\renewcommand\arraystretch{0.7}		
	\begin{equation}\nonumber
		{\bf B} = \begin{bmatrix}
			1 & 1 & 1 & 2\\
			1 & 1 & 0 & 0\\
			1 & 0 & 1 & 0
		\end{bmatrix}
	\end{equation}
	\renewcommand\arraystretch{1}	
	To lift the protograph with $Z = 2$, we first need $2$ copies of the protograph, as shown in Fig.~\ref{fig:protograph_lift}(b). Then, edges of the same type are permuted and parallel edges are ``eliminated'', generating the lifted graph Fig.~\ref{fig:protograph_lift}(c).
\end{example}

\section{Batched Network Codes}\label{section:BNC}

\subsection{Introduction to BNCs}\label{subsection:intro_BNC}
\textit{1) General Description of BNCs: }Different types of BNCs are given distinct names and self-contained descriptions in the literature~\cite{maymounkov2006methods,5191397,overlapChunk,5695118,GammaCodes2012,Tang2012EOC,BATS,Tang2018Lchunked}. Here, we provide a general description of BNCs used for single-source multicasts. Before our description, we first introduce single-source multicast networks.

A single single-source multicast network is represented as a directed graph ${\cal G} = ({\cal V},{\cal E})$, where $\cal V$ and $\cal E$ are the node set and edge set, respectively. $\cal V$ contains a source node $s$, a set $\cal D$ of destination nodes, and some intermediate nodes. Fix a $(K,A)$ code $\cal{P}$, which contains $q^A$ codewords over $\mathbb{F}$ and has the code length $K$. This code is called \textit{precode}.\footnote{We note that in network coding, a precode is not always necessary. The case without a precode is equivalent to using a trivial rate-$1$ precode. Precodes play an important role in the BNCs introduced in Sec.~\ref{section:PBNC}.} At the source node $s$, $A$ packets, each consisting of $T$ symbols over $\mathbb{F}$, are arranged in a matrix ${\bf U} = [u_{i,j}]_{1 \le i \le T, 1 \le j \le A}$, where each column of $\bf U$ represents a packet. Applying the precode $\cal{P}$ to every row of $\bf U$, a matrix ${\bf V} = [v_{i,j}]_{1 \le i \le T, 1 \le j \le K}$ is generated such that each row is a codeword in ${\cal{P}}$. The $A$ packets before precoding and the $K$ coded packets are called \textit{input packets} and \textit{intermediate packets}, respectively. Consider multicasting $\bf V$ to all nodes in $\cal D$. Each member of $\cal D$ aims to recover all input packets. Each edge $e \in {\cal E}$ is modeled as an \textit{erasure channel} and associated with an erasure probability $\epsilon_e$. We assume that all $\epsilon_e$, $e \in {\cal E}$, are independent, and each is in the range $[0,1]$.\footnote{By leveraging the later-defined ${\cal H}_{\cal D}$, it is feasible to extend the analysis in Sec.~\ref{sec:asy} to cases where the range of $\epsilon_e$, $e \in {\cal E}$, is constrained, or even where the channels are correlated.} The set of incoming and outgoing channels of a given node $t \in {\cal V}$ is denoted by ${\rm In}(t)$ and ${\rm Out}(t)$, respectively. We assume ${\cal G}$ does not contain a directed cycle.

Fix a relatively small integer $M$ satisfying $M \ge |{\rm Out}(s)|$. In batched network coding, multicasting $K$ packets generated at the source node is realized by $N$ rounds of multicasting $M$ packets. The $M$ packets to be multicasted in each round is called a \textit{batch}, and $M$ is called the \textit{batch size}. 

A BNC over a network $\cal G$ contains an \textit{outer code} and an \textit{inner code}. At the source node $s$, the outer code generates $N$ batches ${\bf X}_1,{\bf X}_2,\ldots,{\bf X}_N$ for $N$ rounds of multicasting:
	\begin{equation}
		{\bf X}_i = {\bf V} {\bf G}_i,~~i = 1,2,\ldots,N,
	\end{equation}
	where ${\bf G}_i$ is a $K \times M$ matrix over $\mathbb{F}$, called the \textit{generator matrix of the $i$-th batch}. The outer code is a linear code with the generator matrix $[{\bf G}_1,{\bf G}_2,\ldots,{\bf G}_N]$.

The inner code is a linear network code~\cite{1176612} over $\cal G$ with ${\bf X}_i$ being the information to be multicasted in the $i$-th round, which may be described as follows: First, ${\bf X}_i$ is partitioned column-wise into $|{\rm Out}(s)|$ submatrices, which are going to be transmitted through the outgoing edges of $s$. At every node $t \ne s$, if a $P$-column matrix ${\bf Z}$ is received, the encoding function can be written as ${\bf Z}' = {\bf Z} {\bf K}_{i,P}^{(t)}$, where ${\bf K}_{i,P}^{(t)}$ is a $P \times Q$ matrix over $\mathbb{F}$ (encoding at nodes $t \ne s$ is often called \textit{recoding}). Here, $P$ is a variable, but $Q$ is deterministic (may be a function of $t$), and is bounded by $O(M)$. Then, ${\bf Z}'$ is partitioned column-wise into $|{\rm Out}(t)|$ submatrices, which are going to be transmitted through the outgoing edges of $t$. For every node $t$ (including $s$), a method to partition the coded packets into $|{\rm Out}(t)|$ parts can be deterministic, which should be included in the design of the inner code.

Since all operations in the network are linear, the received packets at a node $t \in {\cal D}$ can be written as 
	\begin{equation}\label{eq:linear_equations}
		{\bf Y}_i^{(t)} ={\bf V} {\bf G}_i {\bf H}_{i}^{(t)},~~i = 1,2,\ldots,N,
	\end{equation}
	where ${\bf H}_{i}^{(t)}$ is an $M$-row matrix over $\mathbb{F}$, and is referred to as the \textit{transfer matrix of the $i$-th batch}. A transfer matrix ${\bf H}_{i}^{(t)}$ entirely describes the end-to-end linear transformation from $s$ to $t$, excluding the outer coding by ${\bf G}_i$.

Since erasures occur randomly, a transfer matrix is in fact a random matrix. We assume that all transfer matrices are independent and their ranks follow the same distribution. The \textit{rank distribution} of transfer matrices ${\bf H}_{1}^{(t)},{\bf H}_{2}^{(t)},\ldots,{\bf H}_{N}^{(t)}$ is denoted by a probability vector ${\bf h}^{(t)} = (h_0^{(t)},h_1^{(t)},\ldots,h_M^{(t)})$, where $h_r^{(t)} \triangleq \Pr\{{\rm rk}({\bf H}_{i}^{(t)}) = r\}$. In our network model, ${\bf h}^{(t)}$ is a function of erasure probabilities of channels $\{\epsilon_e: e \in {\cal E}\}$ once all encoding functions are fixed.\footnote{We assume that the erasure probabilities are fixed during the transmission of $N$ batches.} For fixed encoding functions and network topology, the range of ${\bf h}^{(t)}$ is generally not equal to $\{{\bf h} \in [0,1]^{M+1}: \sum_{r=0}^{M} h_r = 1\}$ (the set of all valid distributions), even though we assume that the erasure probabilities are unconstrained. This fact can be verified by considering the point-to-point case (i.e., $|{\cal V}| = 2$ and $|{\cal E}| = 1$): For any erasure probability of the channel, the rank distribution should be a binomial distribution. Let ${\cal H}_t$ denote the set of all possible rank distributions at a destination node $t \in {\cal D}$ and ${\cal H}_{\cal D} \triangleq \bigcup_{t \in {\cal D}} {\cal H}_t$, which depends on  $\cal G$ and all encoding functions. Let ${\cal F}_{\rm outer} = \{{\bf G}_i:i=1,2,\ldots,N\}$ and ${\cal F}_{\rm inner} = \{{\bf K}_{i,P}^{(t)}: t \in {\cal V} \setminus s, i=1,2,\ldots,N, P = 1,2,\ldots \}$ denote the sets of encoding functions of the outer code and the inner code, respectively. We may represent a BNC as a quadruple $({\cal{P}},{\cal F}_{\rm outer}, {\cal F}_{\rm inner},{\cal H}_{\cal D})$.\footnote{We regard ${\cal H}_{\cal D}$ as a part of a BNC for the sake of concise clarity in the later sections.}

\textit{2) Bounded Complexity: } In $\cal G$, let ${\rm In}^* \triangleq \max_{t \in {\cal V} \setminus s}\{|{\rm In}(t)|\}$. We assume ${\rm In}^* = O(1)$. Since an encoding function ${\bf K}_{i,P}^{(t)}$ is a $P \times Q$ matrix with $Q = O(M)$, the variable $P$ can be bounded by $O(M) \times {\rm In}^* = O(M)$. For nodes $t \ne s$, the complexity of recoding a batch is $O(TM^2)$, and the complexity of storing a batch is $O(TM)$. Both complexity are independent of $K$ that may goes to infinity.

In practice, we can bound the average number of non-zero elements in ${\bf G}_i$ by $O(M^2)$ through some specific design. That is, $\lim_{N \to \infty} \frac{\sum_{i=1}^{N}\text{number of non-zero elements in }{\bf G}_i}{N} = O(M^2)$. In this case, the average complexity of encoding a batch at the source node is $O(TM^2)$, independent of $K$. Additionally, we also hope to simplify the decoding through a special design of ${\bf G}_i$, $i = 1,2,\ldots,N$, see more details in Sec.~\ref{subsection:BATS}.

\textit{3) Performance Bound: } Recall that the transmission model of a batch for a node $t$ is ${\bf Y}_i = {\bf V} {\bf G}_i {\bf H}_{i}^{(t)} = {\bf X}_i {\bf H}_{i}^{(t)}$. This model can be regarded as a channel with input ${\bf X}_i$ and output ${\bf Y}_i$. When the node $t$ knows the instance of ${\bf H}_{i}^{(t)}$, the maximal achievable rate (packets per channel use) is~\cite{BATS,LOC2010}:
\begin{equation}\label{eq:network_capacity}
	C^{(t)}({\bf h}^{(t)}) \triangleq \sum_{r=1}^{M} r h_r^{(t)},
\end{equation}
which can be achieved by random linear codes. We call $C^{(t)}({\bf h}^{(t)})$ the capacity of the node $t$. For example, the capacity in Fig.~\ref{fig:achievable_rate} is computed by (\ref{eq:network_capacity}). We can check that $A/C^{(t)}$ is a lower bound on the number of batches for $t$ to recover all input packets with high probability.

To achieve capacity, we need both $A$ and $N$ go to infinity. This implies that $A/C^{(t)}$ may not be a tight bound when $A$, $N$ are not large. The following lemma provides another bound on the optimal performance.

\begin{lemma}\label{lemma:ML_LB}
	Consider a BNC with $A$ inputs packets and $N$ batches. Let ${\bf H}_1,{\bf H}_2,\ldots,{\bf H}_N$ be the transfer matrices for some node $t$. The optimal performance (in terms of error probability) for $t$ is lowered bounded by $\Pr\left\{ \sum_{i=1}^{N}{\rm rk}({\bf H}_i) < A \right\}$.
\end{lemma} 
\begin{IEEEproof}
	The proof follows immediately from the fact that $\sum_{i=1}^{N}{\rm rk}({\bf H}_i) < A$ implies that any decoder cannot uniquely determine the input packets with the equations in (\ref{eq:linear_equations}).
\end{IEEEproof}

We call the bound in Lemma~\ref{lemma:ML_LB} the \textit{maximum-likelihood (ML) lower bound}. For more information on ML decoding of linear codes over an erasure channel, readers can refer to~\cite[Ch. 3.2]{modern_coding_theory}. When ${\rm rk}({\bf H}_1),{\rm rk}({\bf H}_2),\ldots,{\rm rk}({\bf H}_N)$ are independent, the ML lower bound can be computed using convolution.

\subsection{BATS Codes: A Class of Practical BNCs}\label{subsection:BATS}
As a class of BNCs, BATS codes not only enjoy the low encoding and storage complexity as general BNCs, but also achieve low decoding complexity due to an LT-like~\cite{LT,Raptor} design of the outer code. We review the BATS codes proposed in~\cite{BATS}, which are referred to as \textit{standard BATS codes} to distinguish them from the variants proposed in some literature. Unless otherwise specified, we assume that standard BATS codes \textit{do not} use a precode.

\textit{1) Code Construction: } The outer code of a BATS code is a generalized LT code~\cite{LT}, which is constructed as follows: Let ${\bm \Psi} = (\Psi_1,\ldots,\Psi_K)$ be a probability vector, which is called the \textit{degree distribution}. To generate a generator matrix ${\bf G}_i$, we first randomly sample an integer ${\rm dg}_i$ following the distribution $\Pr\{{\rm dg}_i = d\} = \Psi_d$, $d = 1,2,\ldots,K$. Then, sample a set ${\cal I}_i$ from all subsets of $\{1,2,\ldots,K\}$ of size ${\rm dg}_i$ uniformly at random. The generator matrix ${\bf G}_i = [g_{j,k}]$ is formed as
\begin{equation}\nonumber
	g_{j,k} = \begin{cases}
		\text{chosen from }\mathbb{F}\text{ uniformly} & \text{if } j \in {\cal I}_i,\\
		0 & \text{if } j \notin {\cal I}_i.
	\end{cases}
\end{equation} 
Denote by $\tilde{\bf G}_i$ the ${\rm dg}_i \times M$ submatrix of ${\bf G}_i$ formed by the rows with indices in ${\cal I}_i$, and $\tilde{\bf V}_i$ the $T \times {\rm dg}_i$ submatrix of ${\bf V}$ formed by the columns with indices in ${\cal I}_i$. $\tilde{\bf G}_i$ is a \textit{totally random matrix}, i.e., elements of $\tilde{\bf G}_i$ are chosen independently and uniformly at random from $\mathbb{F}$. Encoding by the outer code can be rewritten as follows:
\begin{equation}\label{eq:outer_encoding}
	{\bf X}_i = \tilde{\bf V}_i \tilde{\bf G}_i,~~i = 1,2,\ldots,N.
\end{equation}

The inner code of a BATS code is a random linear network code~\cite{1705002}. That is, each element of an encoding function ${\bf K}_{i,P}^{(t)}$, $t \in {\cal V} \setminus s, i=1,2,\ldots,N, P = 1,2,\ldots$, is chosen from $\mathbb{F}$ uniformly at random. See \cite{BATS,7110514,8606284} for more specific designs of the inner code.

At a node $t \in {\cal D}$, the received packets in the $i$-th batch satisfy the following equation:
\begin{equation}\label{eq:batch_equation}
	{\bf Y}_i^{(t)} = \tilde{\bf V}_i \tilde{\bf G}_i {\bf H}_{i}^{(t)}.
\end{equation}
Eq. (\ref{eq:batch_equation}) is called the \textit{batch equation of the $i$-th batch}.

\textit{2) Decoding Schemes: }Since the outer code is a generalized LT code, low-complexity BP decoding can be used for BATS codes, and is the focus of this paper. In BP decoding, Gaussian elimination is only applied to small linear systems associated with batches (i.e., eq. (\ref{eq:batch_equation})), where each linear system has at most $M$ linearly independent equations; the BP process uses already recovered intermediate packets to advance the undecoded batches through the constraints of the outer code. It is worth noting that when the precode is sparse (e.g., LDPC code), the constraints of the precode can be utilized by BP decoding, allowing the BP process to leverage the constraints of both the outer code and the precode simultaneously. See \cite{BATS,FL_analysis_BATS,Zhu2023LDPCBATS} for more on BP decoding of BATS codes with and without a sparse precode.

Inactivation decoding, which is an efficient ML erasure decoding for linear codes, can also be applied to BATS codes. See \cite{BATS,yang2022bats,Raptor_codes_foundations_and_trends,Raptor} for more on inactivation decoding.

\begin{figure}[!t]
	\centering
	\includegraphics[width=2in]{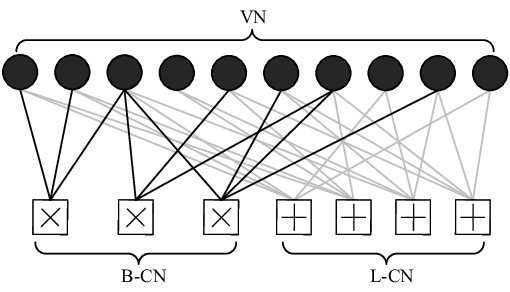}
	\caption{The Tanner graph of a BNC with a $(2,5)$-regular LDPC precode.}
	\label{fig:BATS_Tanner}
\end{figure}

\textit{3) Tanner Graph Representation: }A BATS code can be represented by a Tanner graph. Fig.~\ref{fig:BATS_Tanner} shows the Tanner graph of a standard BATS code with an LDPC precode. For such codes, there are three types of nodes: A VN stands for an intermediate packets, a batch CN (B-CN) corresponds to a constrain defined by a batch equation (\ref{eq:batch_equation}), and an LDPC CN (L-CN) corresponds to a parity-check equation of the LDPC precode. In fact, an L-CN can be regard as a special B-CN with the batch size $1$. In Sec.~\ref{section:PBNC}, we will introduce how to construct such a Tanner graph from a protograph. 

\subsection{Other Examples: Recovery of Several BNCs}\label{subsection:recovery}
Our general description of BNCs recovers many low-complexity network codes, that are essentially BNCs. These codes were described in self-contained ways and were given different names in the literature.

\begin{example}[LDPC-Chunked Codes~\cite{Tang2018Lchunked}]
	An LDPC-chunked code is a BNC with the following specified components: 1) The precode $\cal P$ is a specifically designed LDPC codes with degree-0 VNs. (\textit{Existence of non-coding VNs.}) 2) Fix two integers $L$, $M$, where $L$ is a factor of $K$, and $M \ge L$. ${\bf G}_i$ is a $K \times M$ matrix, and the $((i-1)L+1)$-th to $(iL)$-th rows of ${\bf G}_i$ are totally random rows, while the other rows are all-zero rows. (\textit{All batches have the same degree and are disjoint.}) 3) ${\bf H}_i^{(t)}$ is an $M$-row transfer matrix which describes the end-to-end linear operations.
\end{example}

\begin{example}[Overlapped Chunked Codes~\cite{overlapChunk} and Gamma Codes~\cite{GammaCodes2012}]
	An overlapped chunked code can be described as an LDPC-chunked code with $\cal P$ being a repetition code. This equivalence was clarified in \cite{Tang2018Lchunked}. Similarly, it is straightforward to see that a Gamma code is an LDPC-chunked code with $\cal P$ being a fixed-rate Raptor code. Therefore, overlapped chunked codes and Gamma codes also fall into our BNC framework.
\end{example}

\section{A Protograph Approach to Construct BNCs}\label{section:PBNC}
We propose a protograph approach to construct BNCs. These codes are called P-BNCs. A P-BNC can be viewed as a variation of an LDPC-precoded BATS code (as specified in Fig.~\ref{fig:BATS_Tanner}). In our construction, we assume that rank distributions are given, and only optimize the precode and the outer code. This protograph approach can also be used to construct special classes of BNCs, such as SW-BATS codes~\cite{8629008}, EW-BATS codes~\cite{9664430}, LDPC-chunked codes~\cite{Tang2018Lchunked}, etc., as long as the designed protograph adheres to the structural characteristics of the corresponding class of BNCs. Sec.~\ref{subsection:protograph} describes the structure of a protograph, and then Sec.~\ref{subsection:lifting} introduces how to lift the protograph to obtain a code, as well as how to puncture batches to achieve a higher-rate code.

\subsection{Protograph Structure}\label{subsection:protograph}
First of all, it should be noted that the protograph only specifies the connections between nodes, so the protograph of P-BNCs and the protograph of LDPC codes share no fundamental differences in form, e.g., both can be represented by a \textit{protomatrix}. The main difference between the protograph defined here and the protograph of LDPC codes lies in the specific function of the CNs, that is, what kind of linear constraints they express.

The protograph of a P-BNC consists of two parts: a protograph for the sparse precode and a protograph for the standard BATS code. Dense precodes may not be constructed from protographs and are therefore not discussed here. However, dense precodes can still be applied to P-BNCs (after lifting) to enhance ML decoding. Particularly, a sparse precode (e.g., LDPC code) is always concatenated in P-BNCs, because it not only improves decoding performance but also makes P-BNCs a framework for many existing BNCs, which is briefly introduced in Sec.~\ref{subsection:recovery}. The specific protograph structures of several existing BNCs, such as those proposed in~\cite{Tang2018Lchunked,overlapChunk,GammaCodes2012}, can be found in Appendix~\ref{section:relation}.

The protomatrix of the precode is the same as the general definition of a protomatrix in the LDPC literature. Fix two integers $n_c^{(1)}$ and $n_v$, which are the numbers of L-CNs and VNs, respectively. Let ${\bf B}^{(1)} = [b_{i,j}^{(1)}]_{1 \le i \le n_c^{(1)},1\le j \le n_v}$ be an $n_c^{(1)} \times n_v$ protomatrix for the precode, where every component $b_{i,j}^{(1)}$ is a non-negative integer. Each row corresponds to an L-CN and each column corresponds to a VN. Each L-CN imposes a simple linear constraint such that the sum of neighboring VNs is zero. The role of ${\bf B}^{(1)}$ is to encode $A$ input packets to $K$ intermediate packets. After that, $K$ intermediate packets will be encoded to several batches and transmitted through the network. 

The protomatrix of the standard BATS code is an $n_c^{(2)} \times n_v$ matrix ${\bf B}^{(2)} = [b_{i,j}^{(2)}]_{1 \le i \le n_c^{(2)},1 \le j \le n_v}$, where every component $b_{i,j}^{(2)}$ is a non-negative integer. Similar to an LDPC protomatrix, each row and column of ${\bf B}^{(2)}$ correspond to a B-CN and a VN, respectively. Unlike L-CNs, a B-CN imposes the neighboring VNs to participate in a batch equation, as prescribed by (\ref{eq:batch_equation}). 

Now, the whole protomatrix ${\bf B}$ of a P-BNC can be obtained by juxtaposing matrices ${\bf B}^{(1)}$ and ${\bf B}^{(2)}$, i.e., 
\begin{equation}\nonumber
	{\bf B} = \begin{bmatrix}
		{\bf B}^{(1)}\\
		{\bf B}^{(2)}
	\end{bmatrix} \triangleq [b_{i,j}]_{1 \le i \le n_c, 1 \le j \le n_v},
\end{equation}
where $n_c = n_c^{(1)} + n_c^{(2)}$. Let $d_{c_i} = \sum_{j=1}^{n_v} b_{i,j}$ denote the degree of a type-$i$ CN, where $1 \le i \le n_c$. The $i$-th row of $\bf B$ corresponds to the type-$i$ CN, and the $j$-th column of $\bf B$ corresponds to the type-$j$ VN. 
To avoid ambiguity between the type-$i$ CN, L-CN $i$, and B-CN $i$, we use the following terminology throughout the remainder of the paper: \textit{The $i$-th rows of ${\bf B}^{(1)}$ and ${\bf B}^{(2)}$ are called the L-CN $i$ and the B-CN $i$, respectively. The type-$i$ CN is equivalent to the L-CN $i$ if $i \le n_c^{(1)}$; otherwise, it is equivalent to the B-CN $(i - n_c^{(1)})$.}


For traditional BNCs, the code rate is defined as the ratio of the number of input packets to the number of batches, i.e., $R = A/N$. Similarly, for the protomatrix ${\bf B}$, the design rate is defined as
\begin{equation}\label{eq:design_rate}
	R_{{\bf B}} = \frac{n_v - n_c^{(1)}}{n_c^{(2)}}.
\end{equation}
When the precode constructed from ${\bf B}^{(1)}$ is full rank, the actual rate of the P-BNC is equal to the design rate; otherwise, the actual rate is sightly higher than the design rate. We end this subsection with the following example.

\begin{example}\label{example:protograph}
	Fix batch size $M = 8$, $n_v = 8$, $n_c^{(1)} = 2$, and $n_c^{(2)} = 2$. Consider ${\bf B} = [{\bf B}^{(1)};{\bf B}^{(2)}]$, where
	\renewcommand\arraystretch{0.7}	
	\begin{equation}\nonumber
		{\bf B}^{(1)} = \begin{bmatrix}
			2~1~0~1~0~1~1~1\\
			1~ 1~ 2~ 0~ 2~ 1~ 1~ 1
		\end{bmatrix},~{\bf B}^{(2)} = \begin{bmatrix}
		1~ 1~ 0~ 2~ 1~ 0~ 0~ 1\\
		0~ 1~ 2~ 0~ 1~ 2~ 2~ 2
	\end{bmatrix}.
	\end{equation}
\renewcommand\arraystretch{1}	
	In ${\bf B}^{(2)}$, there must exist some rows with row weight less than or equal to $M$, otherwise BP decoding cannot start. The Tanner graph of such a protomatrix is shown in Fig.~\ref{fig:protograph}.
	\begin{figure}[!t]
		\centering
		\includegraphics[width=2in]{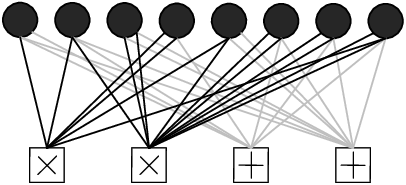}
		\caption{The protograph of $\bf B$ in Example~\ref{example:protograph}.}
		\label{fig:protograph}
	\end{figure}
\end{example}

\subsection{Code Construction From Protograph}\label{subsection:lifting}
Similar to protograph-based LDPC codes, a P-BNC can be obtained by \textit{lifting} the protograph. Although this process is relatively standard, we briefly introduce it here in light of the subtle differences brought about by batched network coding. 

Two-step lifting, similar to that in the LDPC literature~\cite{7045568,5174517,6266764}, is used. The first lifting step is used to remove all parallel edges, i.e., $b_{i,j} > 1$, so the lifting factor $Z_1$ for the first step should satisfy $Z_1 \ge \max_{1 \le i \le n_c, 1 \le j \le n_v} \{b_{i,j}\}$. Let ${\bf T}'$ denote the resulting matrix of size $Z_1 n_c \times Z_1 n_v$. Then, the second lifting step expands ${\bf T}'$ again by a lifting factor $Z_2$ with the \textit{quasi-cyclic constraint} to get the lifted matrix $\bf T$ of size $Z_1 Z_2 n_c \times Z_1 Z_2 n_v$. It implies that every ``1'' in ${\bf T}'$ will be replaced by a $Z_2 \times Z_2$ circulant permutation matrix and every ``0'' in ${\bf T}'$ will be replaced by a $Z_2 \times Z_2$ all-zero matrix. Note that $\bf T$ is a biadjacency matrix that defines the Tanner graph of the code, so it is be necessary to define the constraint for each CN in the Tanner graph to complete the code construction. For the precode part of $\bf T$, non-zero elements in $\mathbb{F}$ need to be labeled to the edges to define a specific LDPC code over $\mathbb{F}$; but for the BATS part of $\bf T$, we generate each batch equation in the same manner as in Sec.~\ref{subsection:BATS}, except that which VNs participate in the batch is specified by $\bf T$.

In general, the progressive edge growth (PEG) algorithm~\cite{PEG} will be used for the two lifting steps. Interestingly, we observe that random lifting (similar to the random lifting introduced in Sec.~\ref{subsection:protograph_codes}, but parallel edges need to be eliminated) for ${\bf B}^{(2)}$, the protograph of the standard BATS code, does not seem to degrade performance. This may be because BATS codes are not as sensitive to short cycles as LDPC codes. A detailed study of the structural weaknesses in BATS codes that lead to BP decoding failure can be found in~\cite{zhu2024TIT}. Roughly speaking, an L-CN is decodable if and only if it has only one unrecovered neighboring VN, while the $i$-th batch can decode about ${\rm rk}(\tilde{\bf G}_i {\bf H}_i^{(t)})$ neighboring VNs (see (\ref{eq:batch_equation})). Thus, we will use the PEG algorithm to lift ${\bf B}^{(1)}$, but randomly lift ${\bf B}^{(2)}$.

Now, we introduce a useful technique for designing high-rate P-BNCs, referred to as \textit{puncturing}. Recall (\ref{eq:design_rate}) for the design rate. If we want to obtain a high-rate P-BNC, the protograph needs to be large or $n_c^{(2)}$ needs to be very small. For example, if $M = 16$ and the capacity of some destination node is $12$ (see (\ref{eq:network_capacity})), even with $n_c^{(1)} = 0$, we need an $n_c^{(2)} \times 12 n_c^{(2)}$ protomatrix ${\bf B}^{(2)}$ to design a code with rate $12$. However, a very small $n_c^{(2)}$ limits the diversity of B-CNs, may leading to bad performance; while, a larger size implies a more complex optimization for the protomatrix. To balance this tradeoff, we can remove some batches from the P-BNC defined by the lifted matrix $\bf T$. This operation can also be described from the perspective of the protograph as follows. Define a length-$n_c^{(2)}$ vector ${\bm \delta} = \left(\delta_1,\delta_2,\ldots,\delta_{n_c^{(2)}}\right)$, called the \textit{puncturing vector}, where $\delta_{i}$ is the puncturing fraction of the B-CN $i$ in ${\bf B}^{(2)}$ (i.e., the type-$(i + n_c^{(1)})$ CN in $\bf B$). For the lifted matrix $\bf T$, $\delta_{i}$ indicates that $\left\lfloor Z_1 Z_2 \delta_{i} \right\rfloor$ rows corresponding to the type-$(i + n_c^{(1)})$ CNs are removed. The modified design rate becomes 
\begin{equation}\nonumber
	R = \frac{n_v - n_c^{(1)}}{n_c^{(2)} - \sum_{i} \delta_i}.
\end{equation}
The selection of punctured rows in $\bf T$ is primarily random, but a small amount of trial and error is also used to avoid some very poor schemes.

\section{Asymptotic Analysis of P-BNCs}\label{sec:asy}
In this section, we develop the asymptotic analysis (i.e., density evolution) of P-BNCs, which is an extension of the tree-based analysis for standard BATS codes in~\cite{Yang2016tree}. In this extension, we need to address several challenges: 1) The tree ensembles corresponding to the proposed protographs need to be defined. 2) The density evolution formulas in~\cite{Yang2016tree} need to be modified to address different decoding behaviors of nodes of different types and the puncturing technique introduced in Sec.~\ref{subsection:lifting}. 3) (Perhaps the most challenging one.) An appropriate optimization objective needs to be defined, which should be applicable to any network topology (e.g., line network, butterfly network~\cite{BATS}) as well as any possible channel conditions (e.g., varying rank distributions). Thus, we can use such an objective to optimize a protograph for any given network.

We address the first and the second challenges in Secs.~\ref{subsection:ensemble} and \ref{subsection:DE}, respectively. For the third challenge, we partially address it in Sec.~\ref{subsection:decoding_threshold}. To be more precise, we propose a suitable optimization objective, called the decoding threshold, for a class of P-BNCs, and we speculate that most P-BNCs should fall into this class.

\subsection{Ensemble and Decoding Neighborhood}\label{subsection:ensemble}
In Sec.~\ref{subsection:lifting}, we introduce the two-step lifting partially with the PEG algorithm. Although this lifting strategy is useful in practice, it is difficult to analyze the asymptotic performance of the codes induced by it. Instead, we will consider the (one-step) random lifting (introduced in Sec.~\ref{subsection:protograph_codes}) with lifting factor $Z$ in the subsequent analysis.\footnote{Any lifting strategy, including the two-step lifting, can be regarded as a specific case of random lifting.}

After random lifting of the protograph $\bf B$, we obtain an $Zn_c \times Zn_v$ matrix $\bf T$. Note that $\bf T$ may contain parallel edges, but for $Z \to \infty$, no parallel edges are in $\bf T$ with high probability. Corresponding to ${\bf B} = [{\bf B}^{(1)};{\bf B}^{(2)}]$, define ${\bf T} = [{\bf T}^{(1)};{\bf T}^{(2)}]$. A P-BNC ensemble is represented by a 5-tuple $({\bf B},{\bf h},M,Z,q)$, where ${\bf B}$ is the protograph, ${\bf h} = (h_0,h_1,\ldots,h_M)$ is the rank distribution of transfer matrices, $M$ is the batch size, $Z$ is the lifting factor, and $q$ is the order of the finite field. Each P-BNC in this ensemble has $Zn_v$ intermediate packets and $Zn_c^{(2)}$ batches. A P-BNC ensemble is defined as follows.
\begin{definition}\label{def:ensemble}
	The $({\bf B},{\bf h},M,Z,q)$ ensemble is a set of P-BNCs such that sampling from this ensemble is equivalent to the following process. First, use random lifting with a lifting factor $Z$ to obtain ${\bf T} \triangleq [{\bf T}^{(1)};{\bf T}^{(2)}]$ from $\bf B$. After labeling each edge in ${\bf T}^{(1)}$ a non-zero element chosen from $\mathbb{F} \setminus \{0\}$ uniformly at random, the precode is obtained. For the $i$-th row in ${\bf T}^{(2)}$, let $\tilde{\bf V}_i = (\tilde{\bf v}_{i,1},\tilde{\bf v}_{i,2},\ldots,\tilde{\bf v}_{i,{\rm dg}_i})$ denote the intermediate packets connected to this row, where $\tilde{\bf v}_{i,j}$ is a column $T$-vector and ${\rm dg}_i$ is the row weight. Sample a generator matrix $\tilde{\bf G}_i$ from $\mathbb{F}^{{\rm dg}_i \times M}$ uniformly at random. Subsequently, sample an $M$-row transfer matrix ${\bf H}_i$ from $\bigcup_{i=1}^{\infty} \mathbb{F}^{M \times i}$ with $\Pr\{{\rm rk}({\bf H}_i)= k\} = h_k$ for $k = 0,1,\ldots,M$. The batch equation of the $i$-the batch is formed as ${\bf Y}_i = \tilde{\bf V}_i \tilde{\bf G}_i {\bf H}_i$. After sampling all batch equations, the P-BNC is determined by the precode and the batch equations.
\end{definition}

\begin{remark}
	One may note that the case where $\bf T$ still includes parallel edges is not explicitly addressed in the above definition. Although we could refine the above definition to encompass this case, we omitted it for simplicity, as the probability of its occurrence approaches zero when $Z \to \infty$.
\end{remark}

\begin{remark}
In the above definition, only one rank distribution is assigned to an ensemble. Recall that ${\cal H}_{\cal D}$ is the set of all possible rank distributions (see Sec.~\ref{subsection:intro_BNC}). To make a P-BNC induced by ${\bf B}$ robust for all rank distributions, we need to consider all ensembles $({\bf B},{\bf h},M,Z,q)$, for ${\bf h} \in {\cal H}_{\cal D}$. This will be elaborated in Sec.~\ref{subsection:decoding_threshold}.
\end{remark}

Define $n_v$ trees, each rooted at a type-$j$ VN, $1 \le j \le n_v$. We define the \textit{height} of a node in the tree as $\lfloor e_{\rm leaf}/2 \rfloor$, where $e_{\rm leaf}$ is the length of the longest downward path to a leaf from this node. If a node $v_i$ has a child $v_j$ in the tree, the edge between $v_i$ and $v_j$ is called $v_i$'s \textit{outgoing edge} or $v_j$'s \textit{incoming edge} (note that the terms ``incoming'' and ``outgoing'' are based on the top-down direction of the tree, which is exactly opposite to the direction of information flow in BP decoding).
\begin{definition}[Tree from node perspective]\label{def:tree}
	Let ${\cal T}_{\ell}(j)$ be a tree of height $\ell$ rooted at a type-$j$ VN. The trivial tree ${\cal T}_{0}(j)$ contains a single node, a fixed VN of type-$j$, serving as the root. ${\cal T}_{\ell}(j)$ is constructed from ${\cal T}_{0}(j)$ recursively by substituting the following ``component trees'':
	\begin{enumerate}
		\item Let $\tilde{\cal L}(j)$ denote a tree rooted at a type-$j$ VN which has $b_{i,j}$ type-$i$ CN children for $i \in \{1,2,\ldots,n_c\}$.
		\item Let ${\cal L}(i,j)$ denote a tree rooted at a type-$j$ VN with a type-$(i,j)$ incoming edge, which has $b_{i',j}$ type-$i'$ CN children for $i' \in \{1,2,\ldots,n_c\} \setminus \{i\}$ and $(b_{i,j}-1)$ type-$i$ CN children.
		\item Let ${\cal R}(i,j)$ denote a tree rooted at a type-$i$ CN with a type-$(i,j)$ incoming edge, which has $b_{i,j'}$ type-$j'$ VN children for $j' \in \{1,2,\ldots,n_v\} \setminus \{j\}$ and $(b_{i,j}-1)$ type-$j$ VN children.
	\end{enumerate}
	${\cal T}_{1}(j)$ is constructed by first substituting the root in ${\cal T}_{0}(j)$ with $\tilde{\cal L}(j)$ and then substituting every type-$i$ leaf CN that has a type-$(i,j)$ incoming edge with ${\cal R}(i,j)$. To construct ${\cal T}_{\ell}(j)$ from ${\cal T}_{\ell-1}(j)$, every type-$j$ leaf VN with a type-$(i,j)$ incoming edge is replaced by ${\cal L}(i,j)$ for all $(i,j) \in \{1,2,\ldots,n_c\} \times \{1,2,\ldots,n_v\}$. Then, every type-$i$ leaf CN with a type-$(i,j)$ incoming edge is replaced by ${\cal R}(i,j)$ for all $(i,j) \in \{1,2,\ldots,n_c\} \times \{1,2,\ldots,n_v\}$.
\end{definition}

\begin{figure}[!t]
	\centering
	\includegraphics[width=3in]{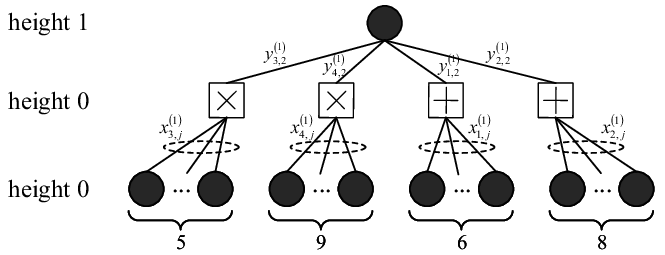}
	\caption{A tree ${\cal T}_{1}(2)$ from the protograph in Example~\ref{example:protograph}. The root is a type-2 VN and the height is 1. The messages used in density evolution are labeled on the edges.}
	\label{fig:tree}
\end{figure}

An example of a tree ${\cal T}_{1}(2)$ is shown in Fig.~\ref{fig:tree}. Following~\cite{910577}, we define the \textit{decoding neighborhood} ${\cal N}_\ell(j)$ of a VN of type-$j$ as the induced subgraph consisting of all nodes reached and edges traversed by paths of length at most $2\ell$. The following theorem shows that for $Z \to \infty$, ${\cal N}_\ell(j)$ is identical to ${\cal T}_{\ell}(j)$ with high probability.

\begin{theorem}\label{theorem:tree_like_neighborhood}
	For an arbitrary but fixed type-$j$ VN, denote by ${\cal N}_\ell^*(j)$ its decoding neighborhood chosen from the $({\bf B},{\bf h},M,Z,q)$ ensemble uniformly at random. For $Z \to \infty$, $\Pr\{{\cal N}_\ell^*(j) = {\cal T}_{\ell}(j)\} \to 1$.
\end{theorem}
\begin{IEEEproof}
	The proof is given in Appendix~\ref{appendix:tree_neighborhood}, following the proof of a tree-like decoding neighborhood for regular LDPC codes~\cite{910577}. However, the probability that a \textit{revealed} edge does not create a loop needs to be reformulated according to the protograph structure. Undoubtedly, this probability approaches 1 as $Z \to \infty$.
\end{IEEEproof}

\subsection{Density Evolution}\label{subsection:DE}
Theorem~\ref{theorem:tree_like_neighborhood} implies that we can analyze the ensemble (Definition~\ref{def:ensemble}) average performance using the tree (Definition~\ref{def:tree}). Consider the BP decoding with $\ell_{\rm max}$ iterations. Then, analyzing the BP decoding is equivalent to analyzing the flow of information from the leaves to the root for a height-$\ell_{\rm max}$ tree. At the $\ell$-th iteration, $1 \le \ell \le \ell_{\rm max}$, we track the erasure probabilities of information flowing from the VNs at height $\ell -1$ to the VNs at height $\ell$.

For a tree defined by Definition~\ref{def:tree}, the following fact holds: At height $\ell$, for all type-$j$ VNs with a type-$(i, j)$ incoming edge (after lifting, there are $Z$ type-$j$ VNs), the \textit{shapes} (e.g., as specified in Fig.~\ref{fig:tree}) of the sub-trees each extended downward from one of the VNs of interest to the leaves are identical. For all CNs at height $\ell$, the situation is the same. This fact implies that we only need to track erasure probabilities based on the edge types (equivalently, the protograph $\bf B$). Let $y_{i,j}^{(\ell)}$ denote the erasure probability of the message transmitted from a CN at height $\ell-1$ to a VN at height $\ell$ via a type-$(i,j)$ edge. Similarly, let $x_{i,j}^{(\ell)}$ denote the erasure probability of the message transmitted from a VN at height $\ell$ to a CN at height $\ell$ via a type-$(i,j)$ edge. If $b_{i,j}=0$, $x_{i,j}^{(\ell)}$ and $y_{i,j}^{(\ell)}$ are fixed as $1$. Specifically, $x_{i,j}^{(0)} = 1$ for all $i,j$. Let $\tilde{\cal N}_v(j)$ be the set of CN types that are connected to a type-$j$ VN. Likewise, let $\tilde{\cal N}_c(i)$ be the set of VN types that are connected to a type-$i$ CN. The density evolution process is described as follows.

For the check-to-variable erasure probability $y_{i,j}^{(\ell)}$, we need to consider two cases: L-CN and B-CN. For an L-CN, its output is non-erased if and only if all of its inputs are non-erased. Thus, $y_{i,j}^{(\ell)}$, $1 \le i \le n_c^{(1)}$, is computed as
\begin{equation}\label{eq:CN_update_1}
	\begin{small}
	\begin{aligned}
	y_{i,j}^{(\ell)} = 1 - \left(\prod_{j' \in {\tilde{\cal N}}_c(i) \setminus \{j\}} \left(1-x_{i,j'}^{(\ell-1)}\right)^{b_{i,j'}}\right) \left(1-x_{i,j}^{(\ell-1)}\right)^{b_{i,j}-1}.
	\end{aligned}
	\end{small}
\end{equation}

Let $\zeta_r^m$ denote the probability that $r$ independently picked totally random vectors of length $m$ over $\mathbb{F}$ are linearly independent, which can be expressed as
\begin{equation}\nonumber
	\zeta_r^m = \begin{cases}
		(1-q^{-m})(1-q^{-m+1})\cdots(1-q^{-m+r-1}) & r > 0,\\
		1 & r = 0.
	\end{cases}
\end{equation}

Consider a B-CN of degree $d$ with $s$ inputs being erased, where $s=0,1,\ldots,d -1$. Let $({\bf Y}_i)_{T \times \cdot} = (\tilde{\bf V}_i)_{T \times d} (\tilde{\bf G}_i)_{d \times M} ({\bf H}_{i})_{M \times \cdot}$ denote the original batch equation of this B-CN, where the subscript in $(\cdot)_{\cdot \times \cdot}$ indicates the size of the matrix. By eliminating the recovered $d-1-s$ packets from the batch equation, the updated equation may be written as $({\bf Y}_i')_{T \times \cdot} = (\tilde{\bf V}_i')_{T \times (s+1)} (\tilde{\bf G}_i')_{(s+1) \times M} ({\bf H}_{i})_{M \times \cdot}$, where $\tilde{\bf V}_i'$ is reduced from $\tilde{\bf V}_i$ by removing $d-1-s$ recovered packets, $\tilde{\bf G}_i'$ is reduced from $\tilde{\bf G}_i$ by removing $d-1-s$ rows corresponding to the recovered packets, and ${\bf Y}_i'$ is adapted from ${\bf Y}_i$. Clearly, the updated equation is decodable if and only if ${\rm rk}(\tilde{\bf G}_i' {\bf H}_{i}) = s+1$. Based on some properties of random matrices (e.g., see Lemma 9 in \cite{zhu2024TIT}), we know that $\left[{\rm rk}(\tilde{\bf G}_i' {\bf H}_{i}) | {\rm rk}({\bf H}_{i}) = r\right]$ is equivalent to the rank of an $(s+1) \times r$ totally random matrix. Thus, we have
\begin{equation}\nonumber
	\Pr\left\{{\rm rk}(\tilde{\bf G}_i' {\bf H}_{i}) = s+1 \bigm| {\rm rk}({\bf H}_{i}) = r\right\} = \zeta_{s+1}^r.
\end{equation}
Considering ${\rm rk}({\bf H}_{i}) = 1,2,\ldots,M$ and $s = 0,1,\ldots,\min\{d-1,r-1\}$ (note that $\zeta_{s+1}^r = 0$ if $s+1 > r$), we can develop the update rule for $y_{i,j}^{(\ell)}$, $n_c^{(1)} + 1 \le i \le n_c$, as follows:
\begin{equation}\label{eq:CN_update_2}
	\begin{aligned}
		y_{i,j}^{(\ell)} &= 1-\sum_{r = 1}^{M} h_r \sum_{s =0}^{\min\{d_{c_i}-1,r-1\}} \Omega(i,j,s,{\bf x}_i^{{(\ell-1)}}) \cdot \zeta_{s+1}^r,
	\end{aligned}
\end{equation}
where $d_{c_i}$ is the degree of a type-$i$ CN, ${\bf x}_i^{{(\ell-1)}}$ is a $n_v$-vector defined as ${\bf x}_i^{{(\ell-1)}}= (x_{i,1}^{(\ell-1)},x_{i,2}^{(\ell-1)},\ldots,x_{i,n_v}^{(\ell-1)})$ (actually only $d_{c_i}$ terms will be used), and $\Omega(i,j,s,{\bf x}_i^{(\ell)})$ is the probability that for the B-CN of interest, $s$ out of $d_{c_i}-1$ incoming messages are erased, and can be formulated as
\begin{equation}\label{eq:Omega}
	\begin{aligned}
		&\Omega(i,j,s,{\bf x}_i^{(\ell)}) \\
		&= \sum_{\substack{k_{j'} = 0,1,\ldots, b_{i,{j'}},\\ j' \in {\tilde{\cal N}}_c(i) \setminus \{j\}}} \sum_{k_{j} = 0}^{b_{i,j}-1} \mathds{1}_{\left\{ \left(\sum_{l \in {\tilde{\cal N}}_c(i)} k_{l}\right) = s \right\}}\\
		&~~~ \cdot \left[\prod_{j' \in {\tilde{\cal N}}_c(i) \setminus \{j\}} \binom{b_{i,j'}}{k_{j'}} \left(x_{i,j'}^{(\ell)}\right)^{k_{j'}} \left(1-x_{i,j'}^{(\ell)}\right)^{b_{i,j'}-k_{j'}}\right]\\
		&~~~ \cdot \binom{b_{i,j}-1}{k_{j}} \left(x_{i,j}^{(\ell)}\right)^{k_j} \left(1-x_{i,j}^{(\ell)}\right)^{b_{i,j}-1-k_{j}},
	\end{aligned}
\end{equation}
where the first summation is over $\prod_{j' \in {\tilde{\cal N}}_c(i) \setminus \{j\}} \{0,1,\ldots, b_{i,{j'}}\}$ (here $\prod$ is the Cartesian product), and $\mathds{1}_{\{\cdot\}}$ is the indicator function.

For a VN, its output is erased if and only if all of its inputs are erased. Consequently, we have the following update rule for variable-to-check erasure probability $x_{i,j}^{(\ell)}$:
\begin{equation}\label{eq:VN_update}
	x_{i,j}^{(\ell)} = \left(\prod_{i' \in \tilde{\cal N}_v(j) \setminus \{i\}} \left( y_{i',j}^{(\ell)} \right)^{b_{i',j}}\right) \left( y_{i,j}^{(\ell)} \right)^{b_{i,j}-1}.
\end{equation}
In particular, if $\ell = \ell_{\rm max}$, the \textit{a posteriori} erasure probability $z_j$ for the root of type-$j$ is updated as
\begin{equation}\label{eq:APP_update}
	z_j^{(\ell_{\rm max})} = \prod_{i \in \tilde{\cal N}_v(j)} \left(y_{i,j}^{(\ell_{\rm max})}\right)^{b_{i,j}}.
\end{equation}

Given $n_v$ types of VNs, the above density evolution process needs to be executed separately for $n_v$ trees with different types of roots. However, similar to the protograph-based extrinsic information transfer (P-EXIT) analysis for LDPC codes~\cite{4411526}, we can efficiently update the information of \textit{all types} of edges based on the protomatrix in each iteration, thereby equating to simultaneously performing density evolution for the $n_v$ different trees. More specifically, in each iteration, edge information is updated for all $(i,j)$ with $b_{i,j} > 0$ using (\ref{eq:CN_update_1}), (\ref{eq:CN_update_2}), or (\ref{eq:VN_update}); in the last iteration, $n_v$ \textit{a posteriori} erasure probabilities $z_1, z_2,\ldots,z_{n_v}$ are updated using (\ref{eq:APP_update}).

Now, we discuss two useful modifications for the above density evolution. One is for puncturing some B-CNs, and another is for simplifying the computation of $\Omega(i,j,s,{\bf x}_i^{(\ell)})$ used in the B-CN update formula (\ref{eq:CN_update_2}). 

First, let us address puncturing. Recall that $\delta_i$ denotes the puncturing fraction of the B-CN $i$ in ${\bf B}^{(2)}$ (i.e., the type-$(i + n_c^{(1)})$ CN in $\bf B$), $1 \le i \le n_c^{(2)}$. Assume that in $\bf T$, $Z \delta_i$ punctured type-$(i + n_c^{(1)})$ CNs are chosen uniformly at random among $Z$ CNs of this type. This random-puncturing assumption implies that each type-$(i + n_c^{(1)},j)$ incoming edge of a B-CN (i.e., the edge between this B-CN and its parent node) has a probability $\delta_i$ to be removed. If we do not change the shape of the tree for analysis, removing an edge is equivalent to set the erasure probability of the message on this edge to one. Using this technique, (\ref{eq:CN_update_2}) can be modified to account for puncturing, which is of the form
\begin{equation}\nonumber
	\begin{aligned}
		y_{i,j}^{(\ell)} =& \delta_{i-n_c^{(1)}} + \left(1-\delta_{i-n_c^{(1)}}\right) \times \\
		& \left(1-\sum_{r = 1}^{M} h_r \sum_{s =0}^{\min\{d_{c_i}-1,r-1\}} \Omega(i,j,s,{\bf x}_i^{{(\ell-1)}}) \cdot \zeta_{s+1}^r \right),
	\end{aligned}
\end{equation}
where $i=n_c^{(1)}+1,n_c^{(1)}+2,\ldots,n_c$.

Then, we provide an approximation to compute $\Omega(i,j,s,{\bf x}_i^{(\ell)})$, i.e., eq. (\ref{eq:Omega}). When $|\tilde{\cal N}_c(i)|$ is large, the computational complexity of $\Omega(i,j,s,{\bf x}_i^{(\ell)})$ becomes prohibitively high. To address this problem, we model $\Omega(i,j,s,{\bf x}_i^{(\ell)})$ as a binomial distribution. This approximation is based on empirical results and some evidence is provided in Appendix~\ref{appendix:approximation}. Let $\tilde\Omega(k;n,p) \triangleq \binom{n}{k} p^k (1-p)^{n-k}$ denote the probability mass function of the binomial distribution. Then, we approximately have
\begin{equation}\label{eq:Omega_approximation}
	\Omega(i,j,s,{\bf x}_i^{(\ell)}) \approx \tilde\Omega(s;d_{c_i}-1,\bar{x}_{i,j}^{(\ell)}),
\end{equation}
where
\begin{equation}\nonumber
	\begin{aligned}
		&\bar{x}_{i,j}^{(\ell)} = \\
		&\begin{cases}
			\frac{1}{d_{c_i} - 1} \left((b_{i,j}-1) x_{i,j}^{(\ell)} + \sum_{j' \in \tilde{\cal N}_c(i) \setminus \{j\}} b_{i,j'} x_{i,j'}^{(\ell)}\right)&d_{c_i} > 1,\\
			1&d_{c_i} = 1.
		\end{cases}
	\end{aligned}
\end{equation}

Using $\tilde\Omega(s;d_{c_i}-1,\bar{x}_{i,j}^{(\ell)})$ as the approximation of $\Omega(i,j,s,{\bf x}_i^{(\ell)})$, the B-CN update formula (\ref{eq:CN_update_2}) becomes
\begin{equation}\label{eq:CN_update_approximate_1}
	\begin{aligned}
		y_{i,j}^{(\ell)} &= 1-\sum_{r = 1}^{M} h_r \sum_{s =0}^{\min\{d_{c_i}-1,r-1\}} \tilde\Omega(s;d_{c_i}-1,\bar{x}_{i,j}^{(\ell-1)}) \cdot \zeta_{s+1}^r.
	\end{aligned}
\end{equation}

The following lemma provides an alternative form of (\ref{eq:CN_update_approximate_1}).
\begin{lemma}\label{lemma:beta_function}
	The approximate B-CN update formula (\ref{eq:CN_update_approximate_1}) can be rewritten with the \textit{regularized incomplete beta function} $I_{x}(a,b)$ as follows:
	\begin{equation}\label{eq:CN_update_approximate_2}
		\begin{aligned}
			&1-\sum_{r = 1}^{M} h_r \sum_{s =0}^{\min\{d-1,r-1\}} {\tilde\Omega}(s;d-1,x) \cdot \zeta_{s+1}^r\\
			&= 1-\sum_{r=1}^{M} I_{1-x}(d-r,r) \sum_{k=r}^{M} \frac{\zeta_r^k}{q^{k-r}} h_{k}
		\end{aligned}
	\end{equation}
	where
	\begin{equation}\nonumber
		I_{x}(d-r,r) = \sum_{s = \max\{0,d-r\}}^{d-1} \binom{d-1}{s} x^s (1-x)^{d-1-s}.
	\end{equation}
\end{lemma}
\begin{IEEEproof}
	Eq. (\ref{eq:CN_update_approximate_2}) can be directly verified through some algebraic calculations (similar to the proof of \cite[Lemma 5.1]{yang2022bats}), but we provide a more intuitive proof in Appendix~\ref{appendix:2}, which gives a simple interpretation for the ``strange'' formula with the regularized incomplete beta function.
\end{IEEEproof}


\subsection{Decoding Threshold}\label{subsection:decoding_threshold}
In this subsection, we propose an optimization objective, called \textit{decoding threshold}, for the optimization of a protograph. In the LDPC literature~\cite{910577}, decoding threshold is generally denoted by a channel parameter, such as erasure probability, noise variance, etc. However, in network communication, the network may have many edges (i.e., channels) and the topology of the network may be arbitrary, making it infeasible to simply borrow the definition of decoding threshold from LDPC codes.

We consider a P-BNC $({\cal{P}},{\cal F}_{\rm outer}, {\cal F}_{\rm inner},{\cal H}_{\cal D})$ used for a multicast network $\cal G$, where ${\cal{P}}$ is the precode, ${\cal F}_{\rm outer}$ and ${\cal F}_{\rm inner}$ are the sets of encoding functions of the outer and inner codes, respectively, and ${\cal H}_{\cal D}$ is the set of all possible rank distributions. For two rank distributions ${\bf h}$ and ${\bf h}'$, we write ${\bf h} \preceq {\bf h}'$ if $\sum_{i=k}^{M} h_i \le \sum_{i=k}^{M} h_i'$ for all $k = 0,1,\ldots,M$. In our code design, we consider that ${\cal F}_{\rm inner}$ and ${\cal H}_{\cal D}$ are fixed. Then, ${\cal{P}}$ and ${\cal F}_{\rm outer}$ will be optimized. Define 
\begin{equation}\nonumber
	{\cal H}_t(C) = \left\{ {\bf h} \in {\cal H}_t: \sum_{r = 1}^{M} r \cdot h_r = C \right\}, ~~{\cal H}_{\cal D}(C) = \bigcup_{t \in {\cal D}} {\cal H}_t(C).
\end{equation}

The decoding threshold will be defined later for a class of P-BNCs (may not apply to all P-BNCs). This class of P-BNCs is defined as follows:
\begin{definition}\label{def:admissible}
	A P-BNC is \textit{admissible} if for any real numbers $C$, $C'$ satisfying $0 \le C' \le C$, and any ${\bf h} \in {\cal H}_{\cal D}(C)$, there exists a ${\bf h}' \in {\cal H}_{\cal D}(C')$ such that ${\bf h}' \preceq {\bf h}$.
\end{definition}

We conjecture that most P-BNCs are admissible, although we could not theoretically establish it. The intuition behind this conjecture is as follows: For a destination node $t$ and a chosen rank distribution ${\bf h} \in {\cal H}_t(C)$, we may obtain a desired $C'$ with $C' \le C$, by increasing the erasure probabilities of some channels on a path from the source node $s$ to $t$. With the degradation of some erasure channels, more packets are likely to be lost during the transmission from $s$ to $t$. Consequently, the probability $\Pr\{{\rm rk}({\bf H}_i^{(t)}) \ge k\}$ (${\bf H}_i^{(t)}$ is a transfer matrix defined in (\ref{eq:batch_equation})) decreases for all $k = 1,2,\ldots,M$, implying the existence of ${\bf h}' \in {\cal H}_t(C')$ satisfying ${\bf h}' \preceq {\bf h}$. We can use the same argument for any $C$, $C'$ ($\le C$), and ${\bf h} \in {\cal H}_{\cal D}(C)$ to show the existence of a proper ${\bf h}'$. We note that when considering line networks, P-BNCs with an inner code as specified in~\cite[VII-A]{BATS} are admissible. The details are given in Appendix~\ref{appendix:admissible}.

Let us first give an intuitive description of the decoding threshold. Consider a protograph $\bf B$ with design rate $R$. Assume that we can find a real number $C$ ($C \ge R$), such that for any ${\bf h} \in {\cal H}_{\cal D}(C)$, the fraction of unrecovered packets (i.e., $z_j^{(\ell_{\rm max})}$, $j = 1,2,\ldots,n_v$, in (\ref{eq:APP_update})) of P-BNCs constructed from $\bf B$ converges to $0$ when $Z \to \infty$. ${\cal H}_{\cal D}(C)$, rather than ${\cal H}_{t}(C)$, should be considered, because all destination nodes $t \in {\cal D}$ need to recover all packets. If such a $C$ is found, we can then naturally infer that the fraction of unrecovered packets should converge to $0$ for any ${\bf h} \in {\cal H}_{\cal D}(C')$ if $C' \ge C$, because the ``channel'' becomes better. This property can be proved using the definition of an \textit{admissible} P-BNC. The minimum of all such $C$ is called the \textit{decoding threshold}, which characterizes the asymptotic performance of P-BNCs constructed from $\bf B$ over a multicast network. We now formally define the decoding threshold as follows:
\begin{definition}[Decoding Threshold]\label{def:threshold}
	For admissible P-BNCs constructed from a protograph, the decoding threshold $C^*$ is defined as 
	\begin{multline}\label{eq:threshold}
		C^* \triangleq \inf \bigg\{ C \in [0,M]: \lim\limits_{\ell_{\rm max} \to \infty} z_j^{(\ell_{\rm max})}({\bf h}) = 0, \\
		\text{ for all }j = 1,2,\ldots,n_v, \text{ and all }{\bf h} \in {\cal H}_{\cal D}(C)  \bigg\},
	\end{multline}
	where $z_j^{(\ell_{\rm max})}({\bf h})$ is the term $z_j^{(\ell_{\rm max})}$ in (\ref{eq:APP_update}) that is calculated by a given rank distribution $\bf h$.
\end{definition}

We now show an important monotonicity property of the decoding threshold.
\begin{theorem}[Monotonicity with respect to channel]\label{theorem:threshold_conotonicity}
	For an admissible P-BNC constructed from a protograph, let
	\begin{multline}\nonumber
		 {\cal C}_0 \triangleq \bigg\{ C \in [0,M]: \lim\limits_{\ell_{\rm max} \to \infty} z_j^{(\ell_{\rm max})}({\bf h}) = 0, \\
		 \text{ for all }j = 1,2,\ldots,n_v, \text{ and all }{\bf h} \in {\cal H}_{\cal D}(C)  \bigg\}.
	\end{multline}
	If $C' \in {\cal C}_0$, then $C \in {\cal C}_0$ for all $C' < C \le M$. 
\end{theorem}
\begin{IEEEproof}
	Let $\bf h$ be a rank distribution in ${\cal H}_{\cal D}(C)$. Since the P-BNC is admissible, there exists ${\bf h}' \in {\cal H}_{\cal D}(C')$ satisfying ${\bf h}' \preceq {\bf h}$, where $C' < C$. Consider the two density evolution processes for $\bf h$ and ${\bf h}'$. Let $x_{i,j}^{(\ell)}({\bf h})$, $y_{i,j}^{(\ell)}({\bf h})$, and $z_{j}^{(\ell_{\rm max})}({\bf h})$ be the erasure probabilities during the density evolution process when the rank distribution is $\bf h$. 
	
	For $\ell = 0$, $x_{i,j}^{(0)}({\bf h}) \le x_{i,j}^{(0)}({\bf h}') = 1$ for all $i,j$. Then, assume that for some $\ell > 0$, $x_{i,j}^{(\ell)}({\bf h}) \le x_{i,j}^{(\ell)}({\bf h}')$ for all $i,j$. Then, noting that (\ref{eq:CN_update_1}) is monotonically increasing in all its arguments, $y_{i,j}^{(\ell+1)}({\bf h}) \le y_{i,j}^{(\ell+1)}({\bf h}')$ holds for all $1 \le i \le n_c^{(1)}$ and $1 \le j \le n_v$. Based on Lemma~\ref{lemma:monotonicity_BCN_1} and Lemma~\ref{lemma:monotonicity_BCN_2} in Appendix~\ref{appendix:3}, whether the exact or approximate B-CN update formula is used, we have $y_{i,j}^{(\ell+1)}({\bf h}) \le y_{i,j}^{(\ell+1)}({\bf h}')$ for all $n_c^{(1)}+1 \le i \le n_c$ and $1 \le j \le n_v$. Thus, we have $y_{i,j}^{(\ell+1)}({\bf h}) \le y_{i,j}^{(\ell+1)}({\bf h}')$ for all $i,j$. Subsequently, observing that (\ref{eq:VN_update}) is monotonically increasing in all its arguments, we have $x_{i,j}^{(\ell+1)}({\bf h}) \le x_{i,j}^{(\ell+1)}({\bf h}')$ for all $i,j$. We conclude by induction that $x_{i,j}^{(\ell_{\rm max})}({\bf h}) \le x_{i,j}^{(\ell_{\rm max})}({\bf h}')$ and $z_{j}^{(\ell_{\rm max})}({\bf h}) \le z_{j}^{(\ell_{\rm max})}({\bf h}')$. 
	
	If $C' \in {\cal C}_0$, we have $z_{j}^{(\ell_{\rm max})}({\bf h}') \to 0$ since ${\bf h}' \in {\cal H}_{\cal D}(C')$. According to $z_{j}^{(\ell_{\rm max})}({\bf h}) \le z_{j}^{(\ell_{\rm max})}({\bf h}')$, we have $z_{j}^{(\ell_{\rm max})}({\bf h}) \to 0$. Since the above argument applies to any ${\bf h} \in {\cal H}_{\cal D}(C)$, $C \in {\cal C}_0$ follows.
\end{IEEEproof}

Theorem~\ref{theorem:threshold_conotonicity} demonstrates that the decoding threshold is well-defined. Thus, we can use the threshold as an objective function to optimize P-BNCs over \textit{any} deterministic single-source multicast network, which will be presented in Sec.~\ref{section:optimization}.

\section{Protograph Optimization and Numerical Results}\label{section:optimization}
In the following Sec.~\ref{subsection:framework}, we first propose an approach to optimize P-BNCs with respect to the proposed decoding threshold. Our goal is to find a specifically structured protograph that achieves a near-optimal decoding threshold. How to find such a protograph from a vast search space is an open problem. We attempt to achieve this goal using randomized algorithms. The efficient search algorithm is not the main focus of this paper. A variety of more sophisticated search algorithms have been employed for design of protograph LDPC codes, such as those based on differential evolution, machine learning, code weaknesses, and others~\cite{10056787,choukroun2024factorgraphoptimizationerrorcorrecting,6266764,7339431,6620513,JPL}. We may draw on these algorithms for reference. In Secs.~\ref{subsection:design_example_1} and \ref{subsection:design_example_2}, we provide numerical results to verify the performance of the optimized P-BNCs. We observe that the optimized P-BNCs significantly outperform existing BNCs, and exhibit robustness under varying channel conditions. The program we used to obtain the numerical results is available on GitHub~\cite{my_github}, and it includes a code construction tool as well as a simulator. Although line networks are considered in the numerical results, our proposed design and analysis can be applied to \textit{any} single-source multicast networks.

\subsection{Protograph Optimization and Discussions}\label{subsection:framework}
We first present the process of optimizing a protograph. Subsequently, we briefly discuss the shortcomings of conventional BATS codes and explain how our design overcomes these limitations. At the end of this subsection, we discuss a simplified but not entirely rigorous method for computing the decoding threshold.

Our approach to optimize a protograph contains four stages: 1) Optimizing the precode. 2) Optimizing the core protomatrix. 3) Optimizing the extension protomatrix. 4) Lifting the protomatrices. Except for stage 1), the process of stages 2) to 4) is similar to the process of optimizing a protograph-based raptor-like LDPC code~\cite{7045568}. As mentioned before, the search algorithms to find a good protograph are not our focus, so we use simple randomized algorithms. One can develop more sophisticated algorithms to improve Algorithms~\ref{alg_opt_core_base_matrix}, \ref{alg_opt_extension_base_matrix}, and \ref{alg_lifting}, which we provide. The four stages for optimizing a protograph are as follows.

\begin{table*}[t]
	\footnotesize
	\caption{Functions for Optimization of Protographs}
	\centering
	\begin{tabular}{|p{4cm}|p{12.5cm}|}
		\hline
		Function & Description \\
		\hline
		\hline
		\texttt{RandMatrix}($m$, $n$, $\bf d$, $s$) & Generate an $m \times n$ random matrix ${\bf B} = [b_{i,j}]_{m \times n}$, where each $b_{i,j} \in \{0,1,\ldots,s\}$ and the $i$-th row has degree $d_i$, $1 \le i \le m$.\\
		\hline
		\texttt{RandRow}($n$, $s$, $D$) & Generate a length-$n$ random row vector ${\bf r} = (r_1,r_2,\ldots,r_n)$, where $r_i \in \{0,1,\ldots,s\}$ and $\sum_{i=1}^{n} r_i \le D$.\\
		\hline
		\texttt{Threshold}($\bf B$, $\bm \delta$) & Calculate the threshold $C^*$ for the protograph $\bf B$ with the puncturing vector $\bm \delta$.\\
		\hline
		\texttt{RandPuncVec}(${\bm \delta}'$) & Generate a puncturing vector $\bm \delta$ such that $\sum_{i} \delta_i = \sum_{i} \delta_i'$. \\
		\hline
		\texttt{PEGLifting}(${\bf B}$, $Z_1$, $Z_2$) & Two-step lifting of ${\bf B}$ using the PEG algorithm with lifting factors $Z_1$ and $Z_2$.\\
		\hline
		\texttt{RandLifting}(${\bf B}$, $\bm \delta$, $Z_1$, $Z_2$) & Two-step random lifting of ${\bf B}$ with lifting factors $Z_1$ and $Z_2$ (but the quasi-cyclic property is preserved). After lifting, ${\bf T}$ is obtained. The returned matrix is obtained by randomly selecting $\lceil (1-\delta_{i}) Z_1 Z_2 \rceil$ rows among $((i-1)Z_1Z_2+1)$-th to $(iZ_1Z_2)$-th rows in ${\bf T}$ for all $i$.\\
		\hline
		\texttt{BPDecodable}(${\bf T}^{(1)}$, ${\bf T}^{(2)}$) & ${\bf T}^{(1)}$ and ${\bf T}^{(2)}$ are the lifted matrices for the sparse precode and the BNC, respectively. Initializing the VNs without any edge (may caused by puncturing) in ${\bf T}^{(2)}$ by erasure probability $1$ and the others by erasure probability $0$, then check if BP decoding of the precode defined by ${\bf T}^{(1)}$ can recover all VNs.\\
		\hline
	\end{tabular}
	\label{table:functions}
\end{table*}

\textit{1) Optimizing the Precode:} First of all, a capacity-approaching protomatrix ${\bf B}^{(1)}$ is designed and will be fixed during the subsequent construction process. Any method used to design protograph LDPC codes can be used to design ${\bf B}^{(1)}$. Additionally, since ${\bf B}^{(1)}$ is only used as a precode, empirical constraints on row and column weights can be discarded in the construction, which are typically used to ensure a low error floor of LDPC codes.

After that, design of ${\bf B}^{(2)}$ can be partitioned into two stages: design of the core part ${\bf B}_{c}^{(2)}$ and design of the extension part ${\bf B}_{e}^{(2)}$, where $[{\bf B}_{c}^{(2)}; {\bf B}_{e}^{(2)}] = {\bf B}^{(2)}$.

\textit{2) Optimizing the Core Protomatrix:} Fixing the precode ${\bf B}^{(1)}$, we design the $m_{c} \times n_v$ core protomatrix ${\bf B}_{c}^{(2)}$. We ensure that the temporary protomatrix $[{\bf B}^{(1)};{\bf B}_{c}^{(2)}]$ has a threshold close to $R(m_{c})$, where $R(j) \triangleq \left({n_v - n_c^{(1)}}\right)/\left({j-\sum_{i=1}^{j} \delta_{i}}\right)$ is the rate of the temporary protograph with $j$ B-CNs, and $\delta_i$ is the puncturing fraction of the B-CN $i$ in ${\bf B}_{c}^{(2)}$.

To optimize ${\bf B}_{c}^{(2)}$ for a given rate $R(m_c)$, we first generate a row degree constraint ${\bf d}_{\rm init} = \left(d_{{\rm init},1},d_{{\rm init},2},\ldots,d_{{\rm init},m_c}\right)$ and a puncturing vector ${\bm \delta}_{\rm init} = \left(\delta_{{\rm init},1},\delta_{{\rm init},2},\ldots,\delta_{{\rm init},m_c}\right)$ empirically, where ${\bm \delta}_{\rm init}$ needs to satisfy the rate. To minimize the threshold of $[{\bf B}^{(1)};{\bf B}_{c}^{(2)}]$, we iteratively optimize the row degree distribution, the column degree distribution, and the puncturing vector by fixing the two out of three objects alternatively. The details are shown in Algorithm~\ref{alg_opt_core_base_matrix} and the functions used in this algorithm are described in Table~\ref{table:functions}. The size of each loop ($i^*,i_{\rm r}^*,i_{\rm c}^*,i_{\rm p}^*$) in Algorithm~\ref{alg_opt_core_base_matrix} has not been explicitly specified, and can be adjusted appropriately to strike a balance between time complexity and the optimality of the result. After this step, we obtain the optimized core protomatrix ${\bf B}_{c}^{(2)}$ and the corresponding optimized puncturing vector ${\bm \delta}_c$.

\begin{algorithm}[t]\label{alg_opt_core_base_matrix}
	\footnotesize
	\SetKwFunction{RandMatrix}{RandMatrix}
	\SetKwFunction{Threshold}{Threshold}
	\SetKwFunction{RandPuncVec}{RandPuncVec}
	
	\caption{Optimizing the Core Protomatrix ${\bf B}_{c}^{(2)}$}
	\KwIn{${\bf d}_{\rm init}$, ${b}_{\rm max}$, ${\bm \delta}_{\rm init}$, ${\bf B}^{(1)}$}
	\KwOut{${\bf B}_{c}^{(2)}$, ${\bm \delta}_c$}
	$C_{\rm min}^* \leftarrow \infty$, ${\bm \delta}_c \leftarrow {\bm \delta}_{\rm init}$\;
	${\bf B}_{c}^{(2)}$ $\leftarrow$ \RandMatrix{$m_c$, $n_v$, ${\bf d}_{\rm init}$, ${{b}}_{\rm max}$}\;
	\For(\tcp*[f]{iterative opt (e.g., ${i}^* \approx 100$)}){$i = 1:{i}^*$}
	{
		Generate $\bf d$, where $d_i = \sum_{j=1}^{n_v} {\bf B}_{\rm c}^{(2)}[i,j]$, $i \in \{1,2,\ldots,m_c\}$\;
		\For(\tcp*[f]{fix row degrees (e.g., $i_{\rm r}^* \approx 1000$)}){$i_{\rm r} = 1:i_{\rm r}^*$}{
			$\tilde{\bf B}$ $\leftarrow$ \RandMatrix{$m_c$, $n_v$, ${\bf d}$, ${{b}}_{\rm max}$}\;
			$C^*$ $\leftarrow$ \Threshold{$[{\bf B}^{(1)};\tilde{\bf B}]$, ${\bm \delta}_c$}\;
			\If{$C^* < C_{\rm min}^*$}{
				${\bf B}_{c}^{(2)} \leftarrow \tilde{\bf B}$, $C_{\rm min}^* \leftarrow C^*$\;
			}
		}
		Generate ${\bf d}'$, where $d_j' = \sum_{i=1}^{m_c} {\bf B}_{\rm c}^{(2)}[i,j]$, $j \in \{1,2,\ldots,n_v\}$\;
		\For(\tcp*[f]{fix col degrees (e.g., $i_{\rm c}^* \approx 1000$)}){$i_{\rm c} = 1:i_{\rm c}^*$}{
			$\tilde{\bf B}^{\mathsf{T}}$ $\leftarrow$ \RandMatrix{$n_v$, $m_c$, ${\bf d}'$, ${{b}}_{\rm max}$}\;
			$C^*$ $\leftarrow$ \Threshold{$[{\bf B}^{(1)};\tilde{\bf B}]$, ${\bm \delta}_c$}\;
			\If{$C^* < C_{\rm min}^*$}{
				${\bf B}_{c}^{(2)} \leftarrow \tilde{\bf B}$, $C_{\rm min}^* \leftarrow C^*$\;
			}
		}
		\For(\tcp*[f]{opt punc vector (e.g., $i_{\rm p}^* \approx 1000$)}){$i_{\rm p} = 1:i_{\rm p}^*$}{
			${\bm \delta}'$ $\leftarrow$ \RandPuncVec{${\bm \delta}_c$}\;
			$C^*$ $\leftarrow$ \Threshold{$[{\bf B}^{(1)};{\bf B}_{c}^{(2)}]$, ${\bm \delta}'$}\;
			\If{$C^* < C_{\rm min}^*$}{
				${\bm \delta}_c \leftarrow {\bm \delta}'$, $C_{\rm min}^* \leftarrow C^*$\;
			}	
		}
	}
\end{algorithm}

\textit{3) Optimizing the Extension Protomatrix:} After optimizing the core protomatrix, ${\bf B}_{\rm c}^{(2)}$ and ${\bm \delta}_c$ is obtained. We then proceed to design the $m_{e} \times n_v$ extension protomatrix ${\bf B}_{e}^{(2)}$. Let ${\bf B}_{{e},s}^{(2)}$ denote the upper $s$ rows of ${\bf B}_{e}^{(2)}$. For each $s$, the temporary protomatrix $[{\bf B}^{(1)};{\bf B}_{c}^{(2)};{\bf B}_{{e},s}^{(2)}]$ is optimized to achieve a threshold close to $R(m_{c} + s)$. In this stage, the puncturing vector ${\bm \delta}_e$ for ${\bf B}_{e}^{(2)}$ is determined empirically and is fixed (the whole puncturing vector is ${\bm \delta} = [{\bm \delta}_c, {\bm \delta}_e]$). When optimizing ${\bf B}_{{e},s}^{(2)}$, we do not adjust the whole puncturing vector as in the optimization of the core protomatrix, because it may degrade the performance of higher-rate parts, i.e., $[{\bf B}^{(1)};{\bf B}_{c}^{(2)};{\bf B}_{{e},s'}^{(2)}]$ with $s' < s$. If the maximum integer in ${\bf B}_{e}^{(2)}$ is $b_{\rm max}'$, then the optimization space is $\{0,1,\ldots,b_{\rm max}'\}^{n_v}$. When both $b_{\rm max}'$ and $n_v$ are small, exhaustive search is feasible. When the optimization space is too large, we can generate a random row with some empirical constraints (e.g., the row weight, the distribution of zeros) and compute the threshold. Repeat this process multiple times to find a local optimum. We use Algorithm~\ref{alg_opt_extension_base_matrix} to realize this procedure. The related functions are described in Table~\ref{table:functions}

\begin{algorithm}[t]\label{alg_opt_extension_base_matrix}
	\footnotesize
	\SetKwFunction{RandRow}{RandRow}
	\SetKwFunction{Threshold}{Threshold}
	
	\caption{Optimizing the Extension Protomatrix ${\bf B}_e^{(2)}$}
	\KwIn{${\bm \delta}$, ${\bf B}^{(1)}$, ${\bf B}_c^{(2)}$, $b_{\rm max}'$}
	\KwOut{${\bf B}_{e}^{(2)}$}
	$C_{\rm min}^* \leftarrow \infty$\;
	${\bf B}_{e,0}^{(2)} \leftarrow []$;\tcp*[f]{assign an empty matrix}\\
	\For(\tcp*[f]{optimize row by row}){$s = 1:m_e$}
	{
		\For(\tcp*[f]{e.g., $i_{\rm r}^* \approx 10000$}){$i_{\rm r} = 1:i_{\rm r}^*$}{
			${\bf r}$ $\leftarrow$ \RandRow{$n_v$, $b_{\rm max}'$, $M$}\;
			$\tilde{\bf B} \leftarrow [{\bf B}_{e,s-1}^{(2)}; {\bf r}]$\;
			$C^*$ $\leftarrow$ \Threshold{$[{\bf B}^{(1)};{\bf B}_c^{(2)};\tilde{\bf B}]$, ${\bm \delta}[1:m_c+s]$}\;
			\If{$C^* < C_{\rm min}^*$}{
				${\bf B}_{e,s}^{(2)} \leftarrow \tilde{\bf B}$, $C_{\rm min}^* \leftarrow C^*$\;
			}
		}
	}
	${\bf B}_{e}^{(2)} \leftarrow {\bf B}_{e,m_e}^{(2)}$\;
\end{algorithm}

\textit{4) Lifting the Protomatrices:} After obtaining ${\bf B}^{(1)}$, ${\bf B}_{c}^{(2)}$, and ${\bf B}_{e}^{(2)}$, we lift them to larger Tanner graphs using Algorithm~\ref{alg_lifting}. The two-step lifting is introduced in Sec.~\ref{subsection:lifting}. Here, we provide some additional explanations on puncturing. Since the puncturing fraction $\delta_i$ may be set to a large value to achieve high rate on a small protograph, e.g., $0.9$, the randomly selected puncturing positions can significantly affect performance, making the resulting code less robust. The function \texttt{BPDecodable} in Algorithm~\ref{alg_lifting} is used to avoid some infeasible puncturing schemes that cannot be recovered by the highest-rate code. However, we cannot guarantee that the code produced by Algorithm~\ref{alg_lifting} will always have good performance. Simulation is still necessary to determine whether the puncturing scheme is satisfactory.

\begin{algorithm}[t]\label{alg_lifting}
	\footnotesize
	\SetKwFunction{RandLifting}{RandLifting}
	\SetKwFunction{PEGLifting}{PEGLifting}
	\SetKwFunction{BPDecodable}{BPDecodable}
	\caption{Lift the Protomatrices}
	\KwIn{$\bm \delta$, ${\bf B}^{(1)}$, ${\bf B}_c^{(2)}$, ${\bf B}_e^{(2)}$, $Z_1$, $Z_2$}
	\KwOut{${\bf T}^{(1)}$,${\bf T}^{(2)}$}
	${\bf T}^{(1)}$ $\leftarrow$ \PEGLifting{${\bf B}^{(1)}$, $Z_1$, $Z_2$}\;
	\While{true}{
		${\bf T}_{c}^{(2)}$ $\leftarrow$ \RandLifting{${\bf B}_c^{(2)}$, ${\bm \delta}[1:m_c]$, $Z_1$, $Z_2$}\;
		\If{\BPDecodable{${\bf T}^{(1)}$, ${\bf T}_{c}^{(2)}$} = true}{
			break\;
		}
	}
	${\bf T}_{e}^{(2)}$ $\leftarrow$ \RandLifting{${\bf B}_e^{(2)}$, ${\bm \delta}[m_c+1:m_c+m_e]$, $Z_1$, $Z_2$}\;
	${\bf T}^{(2)} \leftarrow [{\bf T}_{c}^{(2)}; {\bf T}_{e}^{(2)}]$\;
\end{algorithm}

Some advantages of the protograph approach are discussed as follows.

\textit{Finite-Length Performance: }In most papers on BATS codes, the authors use asymptotic analysis similar to that of LT codes~\cite{LT} to optimize BATS codes~\cite{BATS,Xu2017QUBATS,9664430,8629008}. Such analysis focuses \textit{only} on the degree distribution of B-CNs. However, we observed that such optimized BATS codes (i.e., referred to as conventional BATS codes in this paper) perform poorly at practical lengths (e.g., see Fig.~1 in \cite{FL_analysis_BATS}). Although our protograph approach is still based on asymptotic analysis, our experimental results verified that this approach remains efficient even for relatively short lengths.\footnote{This observation is also supported by a large body of protograph LDPC literature, such as~\cite{Glebov,9797778,6134051,8422961}. In addition, many techniques, which may consider factors such as minimum distance, girth, or simulation results, may be combined with the asymptotic analysis to improve finite-length performance~\cite{8422961,8626506,6134051,8846017}. However, these techniques for further improving finite-length performance are actually separate areas of research and go beyond the scope of this paper.} The reasons why our protograph approach outperforms the approach in~\cite{BATS} at finite lengths may include: First, the asymptotically optimal degree distribution in \cite{BATS} and other BATS literature does not impose a maximum degree, causing a slow convergence from the finite-length performance to the decoding threshold~\cite[Sec. 3.18]{modern_coding_theory}. Second, the ensemble defined by a protograph is a subset of the ensemble merely defined by a degree distribution, so the performance of a random element in the former ensemble might be closer to the decoding threshold, which indicates the average performance of the ensemble. Third, a joint design with an LDPC precode improves the finite-length performance of BATS codes. The improvement caused by LDPC precoding was also observed in~\cite{Zhu2023LDPCBATS,zhu2024TIT}.

\textit{Robustness Against Varying Rank Distributions: }When the erasure probabilities of channels change, the rank distributions at destination nodes also change. For conventional BATS codes, the degree distribution can be optimized for \textit{a set of} rank distributions, as suggested in \cite[Sec. V-C]{BATS}. In this way, conventional BATS codes can also be robust against varying rank distributions. However, the achievable rates of conventional BATS codes optimized by this method are not optimal (e.g., see Fig.~\ref{fig:achievable_rate}, and Table III in~\cite{BATS}). This is because all B-CNs share the same degree distribution. In fact, it is better to address different rank distributions by codes with different rates. Our protograph approach naturally assigns different degree distributions to different B-CNs, allowing the protograph for each design rate to effectively handle the rank distributions it needs to address. This rate-compatible structure is widely adopted in LDPC codes.

We end this subsection with some discussion on the computation of decoding thresholds. In the protograph optimization process, the decoding threshold needs to be computed many times. According to Definition \ref{def:threshold}, each threshold calculation requires iterating over all possible rank distributions\footnote{In our experiments, we use a numerical method to handle the case where $|{\cal H}_{\cal D}(C)|$ may be infinite in decoding threshold calculations. We set the precision of erasure probability and capacity to $\Delta_1$ and $\Delta_2$, respectively. We then traverse all the erasure probabilities in $\{k \Delta_1: k = 0,1,\ldots,\lfloor1/\Delta_1\rfloor\}$, obtaining all the rank distributions. These rank distributions are then categorized into the sets ${\cal H}_{\cal D}(C)$, $C \in \{k \Delta_2: k = 0,1,\ldots,\lfloor M/\Delta_2\rfloor\}$, according to their corresponding capacity. Thus, ${\cal H}_{\cal D}(C)$ becomes a finite set. A theoretical method for optimization with an infinite number of rank distributions may be referenced from~\cite[Ch. 6]{yang2022bats}.}, which is computationally complex. The benefit of this approach is evident: The optimized protograph is robust to variations in rank distributions. One may ask a question: Is it possible to simplify the calculation of decoding thresholds, while keep the robustness? In Sec.~\ref{subsection:design_example_1}, we first strictly follow the definition of a decoding threshold to optimize the protograph. In Sec.~\ref{subsection:design_example_2}, we assume that all erasure probabilities are the same to simplify the calculation of the decoding threshold (in this case, $|{\cal H}_t(C)| = 1$ and $|{\cal H}_{\cal D}(C)| \le |{\cal D}|$, $\forall C$). Simulation results show that the protograph optimized by the simplified method still maintains robustness.

\subsection{Design Example 1}\label{subsection:design_example_1}
In this example, we consider the length-3 line network (i.e., $|{\cal V}| = 4$ and $|{\cal E}| = 3$). The batch size and the order of the finite field are fixed as $M = 8$ and $q = 256$. We use an inner code the same as \cite[Sec. VII-A]{BATS}. Thus, the rank distribution can be calculated by some analytical formulas~\cite{BATS,8606284}. We design a protograph with $n_v = 8$, $m_c = 6$, and $m_e = 6$. We say a network has \textit{homogeneous channels} if all channels have the same erasure probability, and has \textit{heterogeneous channels} otherwise.

First of all, we fix the protograph of the precode as 
\renewcommand\arraystretch{0.7}	
\begin{equation}\label{eq:precode}
	\begin{small}
	\begin{aligned}
		 	{\bf B}^{(1)} = \begin{bmatrix}
			1& 3& 1& 1& 1& 1& 1& 0\\
			1& 3& 2& 0& 1& 0& 0& 1\\
			0& 1& 2& 1& 1& 1& 1& 1
		\end{bmatrix}.
	\end{aligned}
	\end{small}
\end{equation}
\renewcommand\arraystretch{1}	

As the initialization for the optimization, we set the following parameters: ${\bf d}_{\rm init} = (6,8,10,13,15,19)$, ${\bm \delta}_{\rm init} = (0.74,0.86,0.86,0.86,0.92,0.92)$, and $b_{\rm max} = 4$. Following step 2) in Sec.~\ref{subsection:framework}, we obtain the optimized core protomatrix and the optimized puncturing vector for the core protomatrix as follows:
\renewcommand\arraystretch{0.7}	
\begin{equation}\nonumber
	\begin{small}
	\begin{aligned}
			{\bf B}_c^{(2)} = \begin{bmatrix}
			0&1&1&1&1&1&0&2\\
			2&0&3&0&2&0&2&2\\
			1&3&3&0&1&1&1&0\\
			1&3&0&2&0&1&0&3\\
			3&1&3&3&2&4&0&0\\
			4&1&0&4&0&3&2&3
		\end{bmatrix}.
	\end{aligned}
	\end{small}
\end{equation}
\renewcommand\arraystretch{1}	
\begin{equation}\nonumber
	{\bm \delta}_{c} = (0.7292,0.8474,0.8474,0.8339,0.9065,0.9953).
\end{equation}

Then, the puncturing vector ${\bm \delta}_e$ for the extension protomatrix is determined empirically and will not be optimized in step 3) in Sec.~\ref{subsection:framework}. We fix ${\bm \delta}_e$ as
\begin{equation}\nonumber
	\begin{aligned}
		{\bm \delta}_e = (0.88,0.88,0.8,0.8,0.8,0.8)
	\end{aligned}
\end{equation}
and the whole puncturing vector becomes ${\bm \delta} = [{\bm \delta}_c,{\bm \delta}_e]$. Using step 3) in Sec.~\ref{subsection:framework} and setting $b_{\rm max}' = 3$, we obtain the optimized extension protomatrix as follows:
\renewcommand\arraystretch{0.7}	
\begin{equation}\nonumber
	\begin{small}
	\begin{aligned}
		{\bf B}_e^{(2)} = \begin{bmatrix}
			1&3&1&0&1&0&0&1\\
			1&2&1&0&1&0&0&1\\
			1&2&1&0&1&0&0&1\\
			1&1&1&0&1&0&0&1\\
			1&1&1&0&1&0&0&1\\
			0&1&1&0&0&1&0&1
		\end{bmatrix}.
	\end{aligned}
	\end{small}
\end{equation}
\renewcommand\arraystretch{1}
	
Therefore, we obtain the whole protomatrix ${\bf B} = [{\bf B}^{(1)};{\bf B}_c^{(2)};{\bf B}_e^{(2)}]$ with the puncturing vector ${\bm \delta}$. The comparison between the thresholds of this protograph and the code rates is provided in Table~\ref{table:threshold_1}.

\renewcommand\arraystretch{1}	
\begin{table}[t]
	\footnotesize
	\caption{Decoding Thresholds and Code Rates for Design Example 1}
	\centering
	\begin{tabular}{|p{1.7cm}<{\centering}|p{1.5cm}<{\centering}|p{1.5cm}<{\centering}|p{1.5cm}<{\centering}|}
		\hline
		Extended rows &  $C^*$ & $R$ & Gap \\
		\hline
		\hline
		0 &  6.1010 & 5.9524 & 0.1486 \\
		\hline
		1 &  5.5000 & 5.2083 & 0.2917 \\
		\hline
		2 &  4.9760  & 4.6296  & 0.3464\\
		\hline
		3 &  4.3010  & 3.9063  & 0.3947\\
		\hline
		4 &  3.7710  & 3.3784  & 0.3926\\
		\hline
		5 &  3.3560  & 2.9762  & 0.3798\\
		\hline
		6 &  3.0560 & 2.6596 & 0.3964 \\
		\hline
	\end{tabular}\label{table:threshold_1}
\end{table}
\renewcommand\arraystretch{1}	

\begin{figure}[!t]
	\centering
	\includegraphics[width=3in]{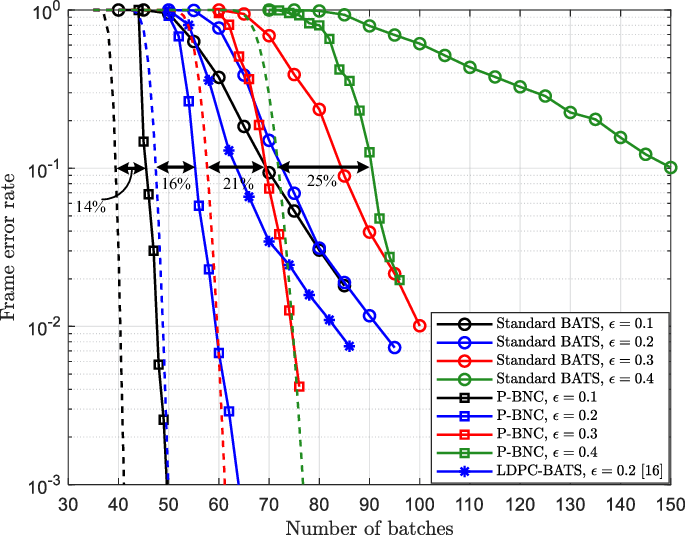}
	\caption{BP performance of the P-BNC in design example 1, the standard BATS code, and the unstructured LDPC-precoded BATS code through the line network of length 3 with homogeneous channels. The dashed curves are the ML lower bounds, distinguished by color to correspond with the solid curves, and the percentages indicated in the figure represent the overhead.}
	\label{fig:line3_fig1}
\end{figure}

\begin{figure}[!t]
	\centering
	\includegraphics[width=3in]{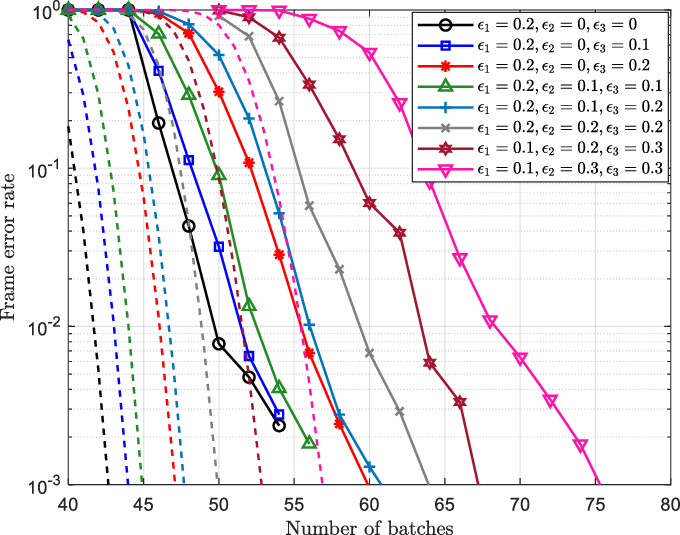}
	\caption{BP performance of the P-BNC in design example 1 through the line network of length 3 with heterogeneous channels. The dashed curves are the ML lower bounds, distinguished by color to correspond with the solid curves.}
	\label{fig:line3_fig2}
\end{figure}

We choose lifting factors $Z_1 = 5$ and $Z_2 = 10$, and then obtain a P-BNC with $A = 250$, $K = 400$, $M = 8$, and $q = 256$. We first compare the performance of the P-BNC, a standard BATS code, and an unstructured LDPC-precoded BATS code~\cite{Zhu2023LDPCBATS} in Fig. \ref{fig:line3_fig1}, where homogeneous channels are considered, and the erasure probability $\epsilon$ is chosen from $\{0.1,0.2,0.3,0.4\}$. The standard BATS code used in Fig. \ref{fig:line3_fig1} is optimized by~\cite[(P3)]{BATS}, which ensures that the achievable rates for all $\epsilon \in \{0.1,0.2,0.3,0.4\}$ are fair in terms of the percentage of capacity. The unstructured LDPC-precoded BATS code is obtained by the finite-length optimization that separately considers the precode and the BATS code. We observe that the P-BNC significantly outperforms the standard BATS code, despite the fact that both are designed based on asymptotic analysis. More surprisingly, the P-BNC performs even better than the unstructured LDPC-precoded BATS code. This supports the importance of the joint design of the precode and BATS code, as well as the role of the protograph in specifying the connectivity. Then, we consider heterogeneous channels and the simulation results are shown in Fig.~\ref{fig:line3_fig2}. We are not surprised by the robustness of this P-BNC under varying erasure probabilities, as we have considered all variations in erasure probabilities in the optimization process. However, we note that the performance curves in Fig.~\ref{fig:line3_fig2} may not be very smooth. In fact, this is understandable because we cannot guarantee that each newly added batch is equally good.

\subsection{Design Example 2}\label{subsection:design_example_2}
In this example, we consider a length-2 line network with homogeneous channels for the optimization of the protograph. When considering homogeneous channels, the capacity is a \textit{decreasing function} of the erasure probability $\epsilon$. Therefore, determining the threshold is equivalent to finding the maximum erasure probability $\epsilon^*$ such that the error probability of the P-BNC converges to zero. The homogeneous-channel model undoubtedly simplifies the calculation of decoding threshold. Later, we will show that the P-BNC designed for homogeneous channels is also robust for heterogeneous channel.

First of all, the protomatrix of the precode is still defined by (\ref{eq:precode}). The batch size and the order of the finite field are fixed as $M = 16$ and $q = 256$. We design a protograph with $n_v = 8$, $m_c = 6$, and $m_e = 8$. Then, following the same steps as those in the design example 1, we obtain

\renewcommand\arraystretch{0.7}	
\begin{equation}\nonumber
	\begin{small}
	\begin{aligned}
		{\bf B}_{c}^{(2)} = \begin{bmatrix}
			2&	5&	2&	1&	2&	1&	0&	3\\
			2&	3&	1&	1&	3&	1&	5&	5\\
			0&	4&	3&	3&	2&	2&	4&	4\\
			3&	2&	3&	0&	1&	0&	1&	4\\
			4&	5&	3&	3&	4&	5&	4&	3\\
			5&	3&	3&	5&	3&	5&	3&	5
		\end{bmatrix},
	\end{aligned}
\end{small}
\end{equation}
\begin{equation}\nonumber
	\begin{small}
	\begin{aligned}
		{\bf B}_{e}^{(2)} = \begin{bmatrix}
			0&	3&	2&	0&	1&	0&	3&	3\\
			0&	3&	2&	0&	3&	0&	1&	3\\
			1&	3&	3&	0&	1&	0&	0&	3\\
			0&	3&	3&	1&	1&	0&	0&	2\\
			0&	2&	0&	0&	3&	1&	0&	3\\
			1&	3&	1&	0&	0&	0&	1&	3\\
			0&	3&	3&	1&	0&	0&	0&	1\\
			0&	2&	2&	0&	2&	0&	0&	2
		\end{bmatrix},
	\end{aligned}
\end{small}
\end{equation}
\renewcommand\arraystretch{1}
and 
\begin{equation}\nonumber
	\begin{aligned}
		{\bm \delta} = & (0.871,0.931,0.927,0.931,0.961,0.961,\\
		&0.94,0.94,0.94,0.94,0.94,0.94,0.94,0.94).
	\end{aligned}
\end{equation}

The comparison between the thresholds of this protograph and the code rates is provided in Table~\ref{table:threshold_2}. Since homogeneous channels are considered, we also provide the erasure probability corresponding to the threshold, denoted by $\epsilon^*$. The results provided in Table~\ref{table:threshold_2} are consistent with Fig.~\ref{fig:achievable_rate}.

\renewcommand\arraystretch{1}	
\begin{table}[t]
	\footnotesize
	\caption{Decoding Thresholds and Code Rates for Design Example 2}
	\centering
	\begin{tabular}{|p{1.7cm}<{\centering}|p{1.2cm}<{\centering}|p{1.2cm}<{\centering}|p{1.2cm}<{\centering}|p{1.2cm}<{\centering}|}
		\hline
		Extended rows & $\epsilon^*$ & $C^*$ & $R$ & Gap \\
		\hline
		\hline
		0 & 0.1904 & 12.0822 & 11.9048 & 0.1774 \\
		\hline
		1 & 0.2588 & 10.8829 & 10.4167 & 0.4662 \\
		\hline
		2 & 0.3193 & 9.8486  & 9.2593  & 0.5893 \\
		\hline
		3 & 0.3730 & 8.9490  & 8.3333  & 0.6157\\
		\hline
		4 & 0.4170 & 8.2240  & 7.5758  & 0.6482\\
		\hline
		5 & 0.4531 & 7.6349  & 6.9444  & 0.6905\\
		\hline
		6 & 0.4863 & 7.0990 & 6.4103 & 0.6887 \\
		\hline
		7 & 0.5176 & 6.5993 & 5.9524 & 0.6469 \\
		\hline
		8 & 0.5459 & 6.1504 & 5.5556 & 0.5948 \\
		\hline
	\end{tabular}\label{table:threshold_2}
\end{table}
\renewcommand\arraystretch{1}	

\begin{figure}[!t]
	\centering
	\includegraphics[width=3in]{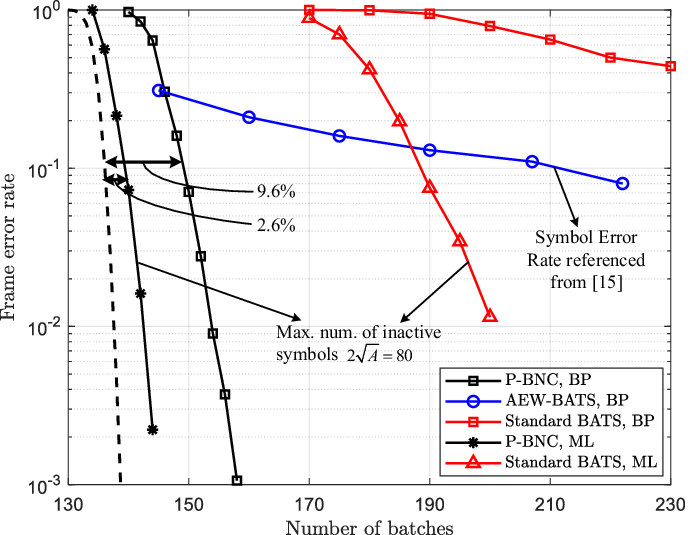}
	\caption{Comparison of the P-BNC in design example 2, the standard BATS code, and the AEW-BATS code~\cite{9664430} through the line network of length 2 with $\epsilon_1 = \epsilon_2 = 0.2$. The ML decoding is implemented by the inactivation decoding of which the maximum number of inactive packets is $2\sqrt{A}$. The dashed curve is the ML lower bound.}
	\label{fig:line2_fig1}
\end{figure}

We lift the protograph using factors $Z_1 = 5$ and $Z_2 = 64$, thus giving a P-BNC with $A =1600$ and $K = 2560$. In Fig.~\ref{fig:line2_fig1}, we compare the BP and ML performance of this P-BNC, the standard BATS code~\cite{BATS} (the same as the one used in Fig.~\ref{fig:achievable_rate}), and the AEW-BATS code~\cite{9664430}, where all codes are with $A = 1600$, $M = 16$, and $q = 256$. Thus, the comparison is made under the same code rate and the same number of input packets. It is not surprising to observe that both the standard BATS code and the AEW-BATS code exhibit \textit{severe error floors} under BP decoding, because their random structures inevitably have some packets that are not well protected without precoding. It is generally believed that this issue can be completely resolved through precoding and ML decoding. Therefore, we also show the ML performance of the standard BATS code with a $q$-ary high-density parity-check (HDPC) precode (adding 40 HDPC packets). We use the inactivation decoding~\cite{Raptor_codes_foundations_and_trends} as the ML decoding algorithm, and the maximum number of inactive packets is limited to $2\sqrt{A}$. However, we see that even though the error floor is eliminated, the ML performance of the standard BATS code still falls significantly short of the ML lower bound. In contrast, under the same complexity constraint, the ML performance of the P-BNC is very close to the ML lower bound, e.g., the overhead is only 2.6\%. Furthermore, even the BP performance of the P-BNC is close to the ML lower bound, and is better than the ML performance of the standard BATS code with a precode.

\begin{figure}[!t]
	\centering
	\includegraphics[width=3in]{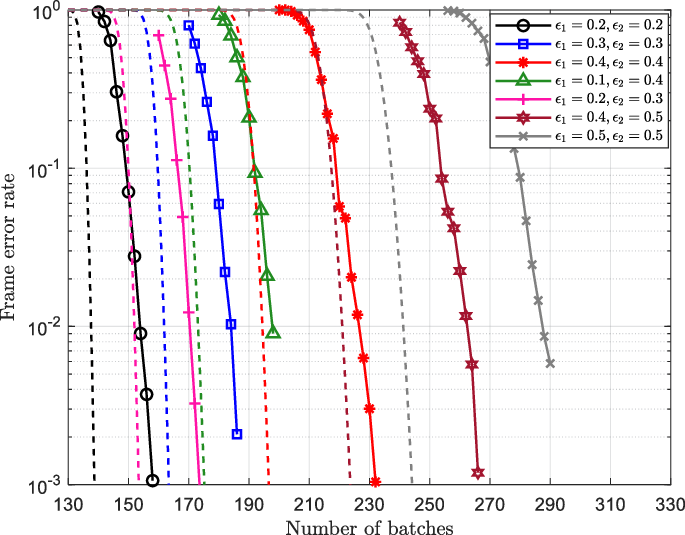}
	\caption{BP performance of the P-BNC in design example 2 through the line network of length 2 with heterogeneous channels. The dashed curves are the ML lower bounds, distinguished by color to correspond with the solid curves.}
	\label{fig:line2_fig2}
\end{figure}

Then, we discuss the robustness of this P-BNC through the length-2 line network with varying erasure probabilities. The simulation results are shown in Fig.~\ref{fig:line2_fig2}. Although this P-BNC is optimized based on the homogeneous-channel model, it also achieves satisfactory performance in many different channels. This observation suggests that designing P-BNCs based on the homogeneous-channel model is sufficient.

\section{Conclusion and Outlook}\label{sec:conclusion}
In this paper, we propose P-BNCs and develop the asymptotic analysis for these codes. Based on the asymptotic analysis and inspired by the design of rate-compatible LDPC codes, we present a framework to design P-BNCs that possess robustness under various channel conditions. P-BNCs achieve excellent BP performance at practical lengths, which is of significant importance for practical applications, and has rarely been observed from the existing BNCs. 

To ensure that the protograph remains compact, we introduce puncturing to delete many batches from the lifted graph. However, the optimal puncturing scheme remains unclear. While the random-puncturing assumption is theoretically acceptable, a specific puncturing scheme heavily impacts the performance in practice. In this study, we select the best puncturing scheme from several randomly generated schemes, but this approach is evidently inefficient. Therefore, specific puncturing schemes may deserve further research.

The computation of decoding threshold following Definition~\ref{def:threshold} is cumbersome. Based on our observation, simplifying the threshold computation by assuming that all channels in a network have the same erasure probability still results in good P-BNCs, which are robust for this network with diverse erasure probabilities. The theoretical basis behind this observation may warrant further investigation.

\appendices
\section{Proof of Theorem \ref{theorem:tree_like_neighborhood}}\label{appendix:tree_neighborhood}
Recall that ${\cal N}_\ell^*(j)$ is a randomly chosen decoding neighborhood for an arbitrary but fixed type-$j$ VN. Let $\texttt{l}_{\rm max}$ and $\texttt{r}_{\rm max}$ be the maximum column weight and row weight of the protomatrix $\bf B$, respectively. Let $V_{\ell}(i,j)$ denote the number of type-$j$ VNs with a type-$(i,j)$ incoming edge in ${\cal N}_\ell^*(j)$. In the same manner, let $C_{\ell}(i,j)$ denote the number of type-$i$ CNs with a type-$(i,j)$ incoming edge in ${\cal N}_\ell^*(j)$. Note that $V_{\ell}(i,j)$ and $C_\ell(i,j)$ are constant random variables.

Fix $\ell$ and let $\ell' < \ell$. Assume that ${\cal N}_{\ell'}^*(j)$ is tree-like, we bound the probability that ${\cal N}_{\ell'+1}^*(j)$ is tree-like. ${\cal N}_{\ell'+1}^*(j)$ is constructed from ${\cal N}_{\ell'}^*(j)$ by first revealing outgoing edges from the leaf VNs, and then from the newly connected leaf CNs. Assume that $k$ additional edges have been revealed from the leaf VNs without creating any loop. Without loss of generality, suppose that the next revealed edge is of type-$(i^*,j^*)$. The probability that the newly revealed type-$(i^*,j^*)$ edge does not create a loop is 
\begin{equation}
	t_c = \frac{\left(Z - \sum_{j=1}^{n_v} C_{\ell'}(i^*,j) - \sum_{j=1}^{n_v} k(i^*,j)\right) b_{i^*,j^*}}{Zb_{i^*,j^*} - C_{\ell'}(i^*,j^*) - k(i^*,j^*) },
\end{equation}
where $k(i,j)$ is the number of type-$(i,j)$ edges among the $k$ already revealed edges. Here, the numerator is the number of type-$(i,j)$ edges that do not produce repeated CNs, and the denominator is the total number of remaining type-$(i,j)$ edges. Noting that $b_{i^*,j^*} > 0$ holds as assuming the revealed edge is of type-$(i^*,j^*)$, $t_c$ can be bounded as
\begin{equation}
	\begin{aligned}
		t_c &\ge \frac{\left(Z - \sum_{j=1}^{n_v} C_{\ell'}(i^*,j) - \sum_{j=1}^{n_v} k(i^*,j)\right) b_{i^*,j^*}}{Zb_{i^*,j^*}}\\
		& \ge 1 - \frac{\sum_{j=1}^{n_v} C_{\ell'}(i^*,j) + \sum_{j=1}^{n_v} k(i^*,j)}{Z}\\
		& \ge 1 - \frac{N_{\ell}}{Z},
	\end{aligned}
\end{equation}
where $N_{\ell}$ is the number of nodes in ${\cal N}_\ell^*(j)$.

Assume that totally $\tilde{k}$ outgoing edges have been revealed from the leaf VNs of ${\cal N}_{\ell'}^*(j)$, among which $\tilde{k}(i,j)$ edges are of type-$(i,j)$. Now consider revealing outgoing edges from the newly connected leaf CNs. Assume that $k$ additional edges have been revealed from the leaf VNs without creating any loop. The next revealed type-$(i^*,j^*)$ edge does not create a loop with the probability
\begin{equation}
	\begin{aligned}
		t_v &=  \frac{\left(Z - \sum_{i=1}^{n_c} V_{\ell'}(i,j^*) - \sum_{i=1}^{n_c} k(i,j^*)\right) b_{i^*,j^*}}{Zb_{i^*,j^*} - V_{\ell'}(i^*,j^*) - k(i^*,j^*) }\\
		&\ge \frac{\left(Z - \sum_{i=1}^{n_c} V_{\ell'}(i,j^*) - \sum_{i=1}^{n_c} k(i,j^*)\right) b_{i^*,j^*}}{Zb_{i^*,j^*}}\\
		& \ge 1 - \frac{\sum_{i=1}^{n_c} V_{\ell'}(i,j^*) + \sum_{i=1}^{n_c} k(i,j^*)}{Z}\\
		& \ge 1 - \frac{N_{\ell}}{Z}.
	\end{aligned}
\end{equation}

Observing that the probability that revealing an edge does not create a loop is bounded by $1 - \frac{N_{\ell}}{Z}$, we have
\begin{equation}
	\begin{aligned}
		\Pr\{{\cal N}_\ell^*(j) = {\cal T}_{\ell}(j)\} & \ge \left(1 - \frac{N_{\ell}}{Z}\right)^{N_{\ell}} = 1- O(Z^{-1}).
	\end{aligned}
\end{equation}

\section{Discussion on the Approximation of $\Omega(i,j,s,{\bf x}_i)$}\label{appendix:approximation}
\begin{figure}[!t]
	\centering
	\includegraphics[width=3.5in]{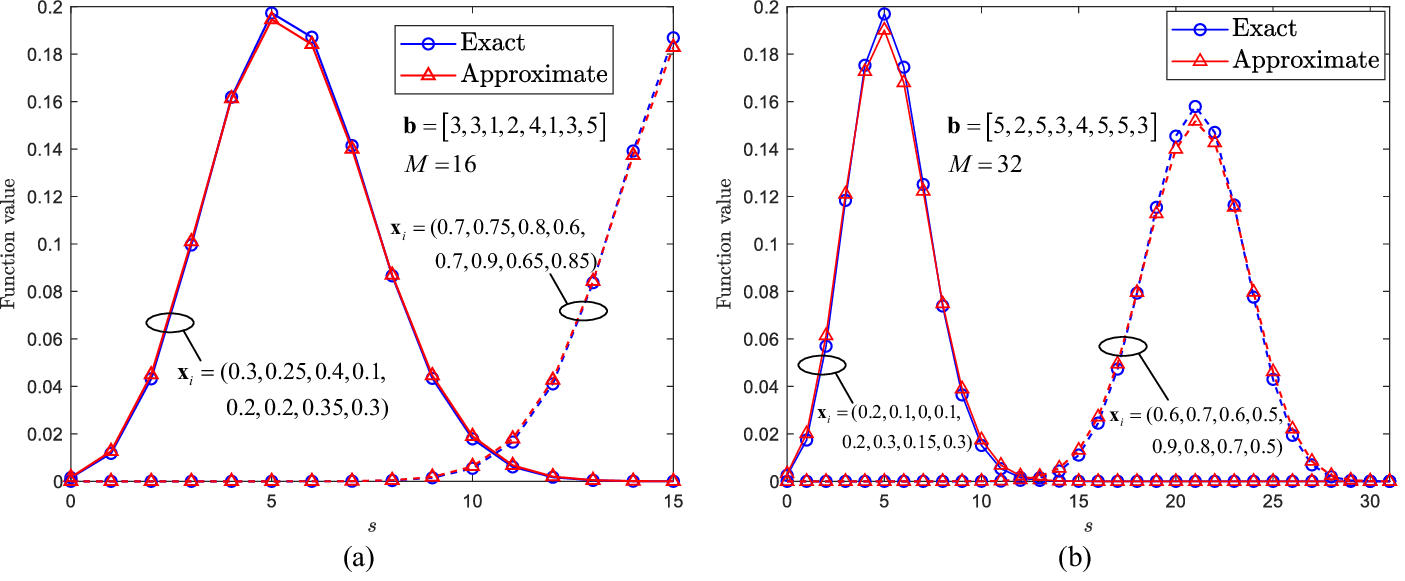}
	\caption{Numerical results on $\Omega(i,j,s,{\bf x}_i)$ and $\tilde\Omega(s;d_{c_i}-1,\bar{x}_{i,j})$.}
	\label{fig:Omega_approx}
\end{figure}

We provide some evidence to demonstrate the accuracy of approximating $\Omega(i,j,s,{\bf x}_i)$ by $\tilde\Omega(s;d_{c_i}-1,\bar{x}_{i,j})$. Assume that ${\bf b} = [b_{i,1}, b_{i,2},\ldots,b_{i,8}]$ is the row of the protograph corresponding to the B-CN $i$. Here, we suppose that $n_v = 8$, consistent with the design examples in Sec.~\ref{section:optimization}. To compare $\Omega(i,j,s,{\bf x}_i)$ and $\tilde\Omega(s;d_{c_i}-1,\bar{x}_{i,j})$, we first randomly set the row ${\bf b}$ as well as the input vector ${\bf x}_i$. Then, $j$ can be fixed as $j=1$ (i.e., consider the output message on the type-$(i,1)$ edge) and test all $s \in \{0,1,\ldots,M-1\}$ (note that only $s \in \{0,1,\ldots,M-1\}$ will be considered in (\ref{eq:Omega})). Fig.~\ref{fig:Omega_approx} shows some numerical results of the comparison, which indicate that the approximation in (\ref{eq:Omega_approximation}) is acceptable.

\section{Proof of Lemma~\ref{lemma:beta_function}}\label{appendix:2}
\begin{IEEEproof}
	This proof proceeds by using two different methods to partition the sample space into mutually exclusive and collectively exhaustive events. Assume that each of $d-1$ incoming messages for the B-CN has a probability $x$ to be erased. Let $E_s$ denote the event that $s$ out of $d-1$ incoming messages are erased. Let $\bf H$ be the transfer matrix associated with the B-CN of interest. Partitioning the sampling space based on the rank of $\bf H$, we can develop
	\begin{equation}\label{eq:BCN_decodable_1}
		\begin{small}
			\begin{aligned}
				&~~~\Pr\{\text{B-CN decodable}\} \\
				&= \sum_{r = 1}^{M} \Pr\{\text{B-CN decodable} \mid {\rm rk}({\bf H})= r\} \Pr\{{\rm rk}({\bf H})= r\}\\
				&= \sum_{r = 1}^{M} h_r \sum_{s=0}^{\min\{d-1,r-1\}}\Pr\left\{ \{\text{B-CN decodable}, E_s\} \mid {\rm rk}({\bf H})= r\right\}\\
				& = \sum_{r = 1}^{M} h_r \sum_{s=0}^{\min\{d-1,r-1\}}\Pr\{\text{B-CN decodable}\mid E_s,{\rm rk}({\bf H})= r\} \\
				&~~~ \times \Pr\{E_s \mid {\rm rk}({\bf H})= r\}\\
				& = \sum_{r = 1}^{M} h_r \sum_{s=0}^{\min\{d-1,r-1\}} \zeta_{s+1}^r {\tilde\Omega}(s;d-1,x)
			\end{aligned}
		\end{small}
	\end{equation} 
	
	Let $D_r$ denote the event that a B-CN is decodable \textit{for the first time} when $E_{r-1}$, i.e., $\Pr\{D_r\} = \Pr\left\{ {\rm rk}\left([{\bf G}^{(1)};{\bf G}^{(r)}] {\bf H}\right) = {\rm rk}\left({\bf G}^{(r)} {\bf H}\right) = r \right\} = \sum_{k=r}^{M} \frac{\zeta_r^k}{q^{k-r}} h_{k}$, where ${\bf G}^{(r)}$ is a $r \times M$ totally random matrix over $\mathbb{F}$. Partitioning the sampling space based on $D_r$, we can write
	\begin{equation}\label{eq:BCN_decodable_2}
		\begin{small}
			\begin{aligned}
				&~~~\Pr\{\text{B-CN decodable}\} \\
				&= \sum_{r = 1}^{M} \Pr\{\text{B-CN decodable} \mid D_r\} \Pr\{D_r\}\\
				&= \sum_{r = 1}^{M} \Pr\left\{\bigcup_{s=0}^{r-1} E_s \mid D_r\right\} \sum_{k=r}^{M} \frac{\zeta_r^k}{q^{k-r}} h_{k}\\
				&= \sum_{r = 1}^{M} \sum_{s=\max\{d-r,0\}}^{d-1}\binom{d-1}{s} (1-x)^s x^{d-1-s} \sum_{k=r}^{M} \frac{\zeta_r^k}{q^{k-r}} h_{k}\\
				& = \sum_{r = 1}^{M} I_{1-x}(d-r,d) \sum_{k=r}^{M} \frac{\zeta_r^k}{q^{k-r}} h_{k}.
			\end{aligned}
		\end{small}
	\end{equation} 
	
	Observing (\ref{eq:BCN_decodable_1}) and (\ref{eq:BCN_decodable_2}), we can claim the equality that $\sum_{r = 1}^{M} h_r \sum_{s=0}^{\min\{d-1,r-1\}} \zeta_{s+1}^r {\tilde\Omega}(s;d-1,x) = \sum_{r = 1}^{M} I_{1-x}(d-r,d) \sum_{k=r}^{M} \frac{\zeta_r^k}{q^{k-r}} h_{k}$. Lemma~\ref{lemma:beta_function} follows.
\end{IEEEproof}

\section{Monotonicity of B-CN Update Functions}\label{appendix:3}
\begin{lemma}\label{lemma:monotonicity_BCN_1}
	Define
	\begin{equation}\nonumber
		\begin{aligned}
			f(x,{\bf h}) = 1-\sum_{r = 1}^{M} h_r \sum_{s =0}^{\min\{d-1,r-1\}} \tilde\Omega(s;d-1,x) \cdot \zeta_{s+1}^r.
		\end{aligned}
	\end{equation}
	1) $f(x,{\bf h})$ is increasing in $x \in (0,1]$. 2) For any ${\bf h}$, ${\bf h}'$ satisfying ${\bf h}' \preceq {\bf h}$, we have $f(x,{\bf h}) \le f(x,{\bf h}')$.
\end{lemma}
\begin{IEEEproof}
	According to Lemma \ref{lemma:beta_function}, $f(x,{\bf h})$ is increasing in $x \in (0,1]$ since $I_{1-x}(d-r,r)$ is decreasing in $x$. Let $\Theta_r = \sum_{s =0}^{\min\{d-1,r-1\}} \tilde\Omega(s;d-1,x) \cdot \zeta_{s+1}^r$ (independent of $\bf h$). Since $\zeta_{s}^r$ is increasing in $r$, $\Theta_r$ is increasing in $r$. We can write
	\begin{equation}\nonumber
		\begin{small}
			\begin{aligned}
				& 1-f(x,{\bf h}) =  \sum_{r = 1}^{M} h_r \Theta_r\\
				&= \Theta_1 \sum_{r = 1}^{M} h_r + (\Theta_2 - \Theta_1) \sum_{r=2}^{M} h_r + \cdots + (\Theta_M - \Theta_{M-1}) \sum_{r=M}^{M} h_r.
			\end{aligned}
		\end{small}
	\end{equation}
	Observing the above equation and noting the constraint ${\bf h}' \preceq {\bf h}$ (i.e., $\sum_{i=k}^{M} h_i' \le \sum_{i=k}^{M} h_i$ for all $k$), $f(x,{\bf h}) \le f(x,{\bf h}')$ is obtained.
\end{IEEEproof}

\begin{lemma}\label{lemma:monotonicity_BCN_2}
	Define
	\begin{equation}\nonumber
		\begin{aligned}
			g_{i,j}({\bf x},{\bf h}) = 1-\sum_{r = 1}^{M} h_r \sum_{s =0}^{\min\{d-1,r-1\}} \Omega(i,j,s,{\bf x}) \cdot \zeta_{s+1}^r,
		\end{aligned}
	\end{equation}
	where $\bf x$ is a length-$n_v$ vector. 1) $g_{i,j}({\bf x},{\bf h})$ is increasing in ${\bf x} \in (0,1]^{n_v}$. 2) For any ${\bf h}$, ${\bf h}'$ satisfying ${\bf h}' \preceq {\bf h}$, we have $g_{i,j}({\bf x},{\bf h}) \le g_{i,j}({\bf x},{\bf h}')$.
\end{lemma}
\begin{IEEEproof}
	Similar to the proof of Lemma \ref{lemma:monotonicity_BCN_1}, 2) can be proved. Next, we proceed to prove 1). Letting $d' \triangleq \min\{d-1,r-1\}$, we can write
	\begin{equation}\label{eq:sum_Omega_zeta}
		\begin{aligned}
			&\sum_{s =0}^{d'} \Omega(i,j,s,{\bf x}) \cdot \zeta_{s+1}^r = \zeta_{d'+1}^r \sum_{s=0}^{d'} \Omega(i,j,s,{\bf x}) +\left(\zeta_{d'}^r-\zeta_{d'+1}^r\right)\\ 
			&\cdot \sum_{s=0}^{d'-1} \Omega(i,j,s,{\bf x}) + \cdots + \left(\zeta_{1}^r-\zeta_{2}^r\right) \sum_{s=0}^{0} \Omega(i,j,s,{\bf x}).
		\end{aligned}
	\end{equation}
	Note that $\zeta_{s}^r-\zeta_{s+1}^r > 0$ holds for $1 \le s \le d'$. Moreover, $\sum_{s =0}^{d'} \Omega(i,j,s,{\bf x})$ is the probability that the number of erased incoming messages is less than or equal to $d'$, implying that (\ref{eq:sum_Omega_zeta}) is decreasing in ${\bf x}$. Thus,  $g_{i,j}({\bf x},{\bf h})$ is increasing in ${\bf x}$.
\end{IEEEproof}

\begin{figure*}[t]
	\normalsize
	\begin{equation}\label{eq:rk_H_k}
		\begin{small}
			\begin{aligned}
				&\Pr\left\{{\rm rk}({\bf H}^{(k)}) \ge t\right\} = \Pr\left\{{\rm rk}({\bf H}^{(k-1)}{\bf T}_k {\bf E}_k) \ge t, {\rm rk}({\bf E}_k) \ge t \right\}\\
				& = \sum_{r=t}^{M} \Pr\left\{{\rm rk}({\bf H}^{(k-1)}{\bf T}_k {\bf E}_k) \ge t \mid {\rm rk}({\bf E}_k) = r\right\} \Pr\left\{{\rm rk}({\bf E}_k) =r\right\}\\
				& = \Theta_{t|t} \sum_{r=t}^{M} \Pr\left\{{\rm rk}({\bf E}_k) =r\right\} + \left(\Theta_{t|t+1} -\Theta_{t|t} \right)\sum_{r=t+1}^{M} \Pr\left\{{\rm rk}({\bf E}_k) =r\right\}  +\cdots + \left(\Theta_{t|M} -\Theta_{t|M-1} \right)\sum_{r=M}^{M} \Pr\left\{{\rm rk}({\bf E}_k) =r\right\}
			\end{aligned}
		\end{small}
	\end{equation}
	\begin{equation}\label{eq:rk_H_k_prime}
		\begin{small}
			\begin{aligned}
				&\Pr\left\{{\rm rk}({\bf H}^{(k')}) \ge t\right\} = \Pr\left\{{\rm rk}({\bf H}^{(k'-1)}{\bf T}_{k'} {\bf E}_{k'}) \ge t, {\rm rk}({\bf H}^{(k'-1)}) \ge t \right\}\\
				& = \sum_{r=t}^{M} \Pr\left\{{\rm rk}({\bf H}^{(k'-1)}{\bf T}_{k'} {\bf E}_{k'}) \ge t \mid {\rm rk}({\bf H}^{(k'-1)}) = r\right\} \Pr\left\{{\rm rk}({\bf H}^{(k'-1)}) =r\right\}\\			
				& = \Upsilon_{t|t} \sum_{r=t}^{M} \Pr\left\{{\rm rk}({\bf H}^{(k'-1)}) =r\right\} + \left(\Upsilon_{t|t+1} -\Upsilon_{t|t} \right)\sum_{r=t+1}^{M} \Pr\left\{{\rm rk}({\bf H}^{(k'-1)}) =r\right\}  +\cdots + \left(\Upsilon_{t|M} -\Upsilon_{t|M-1} \right)\sum_{r=M}^{M} \Pr\left\{{\rm rk}({\bf H}^{(k'-1)}) =r\right\}
			\end{aligned}
		\end{small}
	\end{equation}
	\hrulefill
	\vspace*{4pt}
\end{figure*}
\section{Admissible BNCs For Line Networks}\label{appendix:admissible}
We find that BNCs, whose inner code is a random linear network code (RLNC) as specified in~\cite[VII-A]{BATS}, are admissible when considering line networks. Consider a line network ${\cal G} = ({\cal V},{\cal E})$, where $|{\cal E}| = E$ and $|{\cal V}| = E+1$. Let ${\bm \epsilon} = (\epsilon_1,\epsilon_2,\ldots,\epsilon_E)$, where $\epsilon_i$ is the erasure probability of the $i$-th channel. The main result of this section is given as follows.

\begin{proposition}\label{prop:line_network}
	For a line network, BNCs with the inner code being an RLNC as specified in~\cite[VII-A]{BATS} are admissible.
\end{proposition}
\begin{IEEEproof}
Let ${\bf H}$ be the transfer matrix of some batch for the destination node of the line network (the subscript of $\bf H$ that stands for the index of the batch is omitted). Recall the following definitions: $h_r = \Pr\{{\rm rk}({\bf H}) = r\}$, $\sum_{r=t}^{M} h_r = \Pr\{{\rm rk}({\bf H}) \ge t\}$, and $C({\bf h}) = \sum_{r=1}^{M} rh_r$.

We first review the RLNC over a line network in \cite[VII-A]{BATS}. Assume that the line network has $E+1$ nodes $v_0,v_1,\ldots,v_E$ and $E$ channels $(v_0,v_1),(v_1,v_2),\ldots,(v_{E-1},v_E)$. The $i$-the channel $(v_{i-1},v_i)$ has an erasure probability $\epsilon_i$. Let ${\bm \epsilon} = (\epsilon_1,\epsilon_2,\ldots,\epsilon_E)$. For the first channel, $M$ packets encoded by the outer code will be transmitted, and no RLNC is applied. For the $i$-th channel, $2 \le i \le E$, let ${\bf T}_{i}$ be an $M \times M$ linear transformation matrix (also known as coefficient matrix) of the RLNC, which is a totally random matrix. Let ${\bf E}_i$ denote an $M \times M$ random diagonal matrix, where a diagonal component is 0 with probability $\epsilon_i$ and is 1 with probability $1-\epsilon_i$. So, the transfer matrix ${\bf H}^{(i)}$ for the node $v_i$ can be expressed as
\begin{equation}
	{\bf H}^{(i)} = {\bf H}^{(i-1)} {\bf T}_i {\bf E}_i,~i = 2,3,\ldots,E+1,
\end{equation}
where ${\bf H}^{(1)} = {\bf E}_1$. Let ${\bf h}^{(i)} = (h_0^{(i)},h_1^{(i)},\ldots,h_M^{(i)})$, where $h_r^{(i)} \triangleq \Pr\{{\rm rk}({\bf H}^{(i)}) = r\}$, which can be recursively calculated as
\begin{equation}\label{eq:rank_distribution_recursion}
	\begin{small}
		\begin{aligned}
			h_r^{(i)}  = \sum_{s=r}^{M} \sum_{j=r}^{M} h_s^{(i-1)} \binom{M}{j}			(1-\epsilon_i)^j \epsilon_i^{M-j} \frac{\zeta_r^s \zeta_r^j}{\zeta_r^r q^{(s-r)(j-r)}}.
		\end{aligned}
	\end{small}
\end{equation}

Based on Definition~\ref{def:admissible}, a proper ${\bf h}' \in {\cal H}_{\cal D}(C')$ needs to be found for every $C' < C$. We use two steps to prove Proposition~\ref{prop:line_network}. First, we will show that for a ${\bf h} \in {\cal H}_{\cal D}(C)$, there exists a ${\bf h}'$ satisfying ${\bf h}' \preceq {\bf h}$. Second, we show that such a ${\bf h}'$ can be found in ${\cal H}_{\cal D}(C')$. Noting that the choice of ${\bf h}$ and $C'$ are arbitrary, Proposition~\ref{prop:line_network} can be proved.

\textit{First Step: } Consider increasing the erasure probability of the $k$-th channel $(v_{k-1},v_k)$, say $\epsilon_k' \ge \epsilon_k$. We want to show that $\Pr\left\{{\rm rk}(\tilde{\bf H}^{(E)}) \ge t\right\} \le \Pr\left\{{\rm rk}({\bf H}^{(E)}) \ge t\right\}$ holds for all $t = 0,1,\ldots,M$, where $\tilde{\bf H}^{(i)}$ is the transfer matrix for $v_i$ in the modified network. Clearly, $\tilde{\bf H}^{(i)}$ and ${\bf H}^{(i)}$, $i = 1,2,\ldots,k-1$, are i.i.d. We first consider ${\bf H}^{(k)}$ ($\tilde{\bf H}^{(k)}$) for $v_k$. Define
\begin{equation}\nonumber
	\Theta_{t|r} = \Pr\left\{{\rm rk}({\bf H}^{(k-1)}{\bf T}_k {\bf E}_k) \ge t \mid {\rm rk}({\bf E}_k) = r\right\}.
\end{equation} 
Then, the tail distribution of ${\rm rk}({\bf H}^{(k)})$ can be expressed as (\ref{eq:rk_H_k}), as shown at the top of this page. Note that $\Theta_{t|r}$ is independent of the erasure probability $\epsilon_k$ (due to the given ${\rm rk}({\bf E}_k)$), and is increasing in $r$. Since ${\rm rk}({\bf E}_k)$ follows the binomial distribution $\tilde\Omega(\cdot;M,1-\epsilon_k)$, we can check that $\sum_{r=j}^{M} \Pr\left\{{\rm rk}({\bf E}_k) =r\right\}$ is decreasing in $\epsilon_k$ for all $j=0,1,\ldots,M$. Thus, increasing $\epsilon_k$ to $\epsilon_k'$ leads to $\Pr\left\{{\rm rk}({\bf H}^{(k)}) \ge t\right\} \ge \Pr\left\{{\rm rk}(\tilde{\bf H}^{(k)}) \ge t\right\}$ for all $t=0,1,\ldots,M$. 

We proceed to consider ${\bf H}^{(k')}$ ($\tilde{\bf H}^{(k')}$) for $k' > k$. Since ${\bf T}_{k'}$ is a totally random matrix, we can check that the following two conditional probabilities are equivalent: 
\begin{equation}\nonumber
	\begin{aligned}
		&\Pr\left\{{\rm rk}({\bf H}^{(k'-1)}{\bf T}_{k'} {\bf E}_{k'}) \ge t \mid {\rm rk}({\bf H}^{(k'-1)}) = r\right\} \\
		&= \Pr\left\{{\rm rk}(\tilde{\bf H}^{(k'-1)}{\bf T}_{k'} {\bf E}_{k'}) \ge t \mid {\rm rk}(\tilde{\bf H}^{(k'-1)}) = r\right\} \\
		&\triangleq \Upsilon_{t|r}.
	\end{aligned}
\end{equation}
This is because ${\rm rk}({\bf H}^{(k'-1)}{\bf T}_{k'} {\bf E}_{k'})$ is independent of any structure of ${\bf H}^{(k'-1)}$ if ${\rm rk}({\bf H}^{(k'-1)})$ is given. According to the definition, we know that $\Upsilon_{t|r}$ is increasing in $r$. For $k' > k$, we can develop the tail distribution of ${\rm rk}({\bf H}^{(k')})$ as (\ref{eq:rk_H_k_prime}), as shown at the top of this page. For $\tilde{\bf H}^{(k')}$, the formula is the same as (\ref{eq:rk_H_k_prime}) except that ${\bf H}^{(k'-1)}$ should be replaced by $\tilde{\bf H}^{(k'-1)}$. We are ready to prove $\Pr\left\{{\rm rk}({\bf H}^{(k')}) \ge t\right\} \ge \Pr\left\{{\rm rk}(\tilde{\bf H}^{(k')}) \ge t\right\}$ for all $k' = k+1,k+2,\ldots,E$ and $t=0,1,\ldots,M$ by induction. The base case $\Pr\left\{{\rm rk}({\bf H}^{(k)}) \ge t\right\} \ge \Pr\left\{{\rm rk}(\tilde{\bf H}^{(k)}) \ge t\right\}$ for all $t$ has been proved earlier. Suppose that $\Pr\left\{{\rm rk}({\bf H}^{(k'-1)}) \ge t\right\} \ge \Pr\left\{{\rm rk}(\tilde{\bf H}^{(k'-1)}) \ge t\right\}$ holds for all $t$. Then, from (\ref{eq:rk_H_k_prime}), it is evident that $\Pr\left\{{\rm rk}({\bf H}^{(k')}) \ge t\right\} \ge \Pr\left\{{\rm rk}(\tilde{\bf H}^{(k')}) \ge t\right\}$ holds.

At this point, a ${\bf h}'$ satisfying ${\bf h}' \preceq {\bf h}$ is found.

\textit{Second Step: }It remains to show that a proper ${\bf h}'$ can be found in ${\cal H}_{\cal D}(C')$, $\forall C' < C({\bf h})$. Recall the assumption in Sec.~\ref{subsection:intro_BNC}: All $\epsilon_i$ are independent and are in the range $[0,1]$. Under this assumption, any ${\bm \epsilon} \in [0,1]^E$ is achievable. For clarity, we adopt the following two notations:
\begin{itemize}
	\item $h_i({\bm \epsilon})$ represents the term $h_i$ in a rank distribution $\bf h$ that is induced by the erasure probability vector ${\bm \epsilon}$.
	\item $C({\bm \epsilon}) \triangleq C({\bf h})$, where ${\bf h} = (h_0({\bm \epsilon}),h_1({\bm \epsilon}),\ldots,h_M({\bm \epsilon}))$. 
\end{itemize}

Note that the capacity can be expressed as 
\begin{equation}\nonumber
	\begin{aligned}
		C({\bm \epsilon}) &= \sum_{i=1}^{M} ih_i({\bm \epsilon}) \\
		&= \sum_{i=1}^{M} h_i({\bm \epsilon}) + \sum_{i=2}^{M} h_i({\bm \epsilon}) + \cdots + \sum_{i=M}^{M} h_i({\bm \epsilon}).
	\end{aligned}
\end{equation}
Since $\sum_{i=k}^{M} h_i({\bm \epsilon})$ has been proved to be decreasing in $\bm \epsilon$, we can claim that $C({\bm \epsilon})$ is also decreasing in ${\bm \epsilon}$. Additionally, $C({\bm \epsilon})$ is a continuous function on its domain $[0,1]^E$ because all $h_i({\bm \epsilon})$ (obtained by expanding (\ref{eq:rank_distribution_recursion})) are continuous. Based on the above properties of $C({\bm \epsilon})$ and noting that $C({\bf 1}) = 0$, we can find a ${\bm \epsilon}' \in \prod_{i=1}^{E} [\epsilon_i,1]$ (here, $\prod$ denotes the Cartesian product) such that $C({\bm \epsilon}') = C'$ for any $0 \le C' < C({\bm \epsilon})$. By the assumption mentioned before, this ${\bm \epsilon}'$ is achievable. Thus, we find a ${\bf h}'$ induced by ${\bm \epsilon}'$, which is in ${\cal H}_{\cal D}(C')$.
\end{IEEEproof}

\section{Protograph Generalization of Several BNCs}\label{section:relation}
In Sec.~\ref{section:PBNC}, we introduce P-BNCs as a variant of LDPC-precoded BATS codes based on the protograph. Here, we show that LDPC-chunked (L-chunked) codes \cite{Tang2018Lchunked}, overlapped chunked codes \cite{overlapChunk}, and Gamma codes \cite{GammaCodes2012} can also be constructed from protographs, and their protograph versions are special P-BNCs.

\subsection{LDPC-Chunked Codes}\label{subsection:Lchunked}
LDPC-chunked codes are briefly introduced in Sec.~\ref{subsection:recovery}. Here we review more on LDPC-chunked codes, and then provide their protograph structure. In LDPC-chunked codes, the LDPC precode is specifically designed to have some degree-0 VNs, referred to as \textit{non-coding VNs} in~\cite{Tang2018Lchunked}. After precoding, the intermediate packets are divided into several \textit{disjoint chunks} (which is equivalent to disjoint batches), each consisting of $L$ packets. The integer $L$ is called the chunk size (which is essentially the degree of each batch). Both the outer code and the inner code apply to chunks. Fix $N$ integers $M_1,M_2,\ldots,M_N$. ${\bf G}_i$ is a $K \times M_i$ matrix, and the $((i-1)L+1)$-th to $(iL)$-th rows of ${\bf G}_i$ are totally random rows, while the other rows are all-zero rows. In \cite{Tang2018Lchunked}, $M_1,M_2,\ldots,M_N$ can be different. However, for simplicity of expression, we assume that $M_1 = M_2 = \cdots = M_N \triangleq M$. The extension to the construction with different $M_i$ is straightforward. The transfer matrix ${\bf H}_i^{(t)}$ has the same operational meaning as in Sec.~\ref{subsection:intro_BNC}.

Given $L$, assume that the number $n_v$ of VNs is a multiple of $L$. Then, there are $n_v/L$ chunks in the protograph. After lifting with a lifting factor $Z$, we will obtain $Zn_v/L$ chunks. ${\bf B}^{(1)}$ is still an LDPC protomatrix, but some columns in ${\bf B}^{(1)}$ can be all zeros, corresponding to the non-coding VNs. The form of ${\bf B}^{(2)}$ is fixed as a block diagonal matrix, namely
\renewcommand\arraystretch{0.7}	
\begin{equation}\label{eq:L_chunk_B2}
	{\bf B}^{(2)} = \begin{bmatrix}
		{\bf 1}_{1 \times L} & & & \\
		& {\bf 1}_{1 \times L} & & \\
		& & \ddots & \\
		& & & {\bf 1}_{1 \times L}
	\end{bmatrix}_{\frac{n_v}{L} \times \frac{n_v}{L}},
\end{equation}
\renewcommand\arraystretch{1}	
where ${\bf 1}_{m \times n}$ is an $m \times n$ all-one matrix.

Here, $n_c^{(2)} = {n_v}/{L}$ is implied. Since the column weight of ${\bf B}^{(2)}$ is one, no intermediate packets will be involved in more than one chunks after lifting the protograph, which preserves the key property of LDPC-chunked codes.

\begin{example}
	Consider the LDPC-chunked code with $L = 4$ in \cite[Fig.~3]{Tang2018Lchunked}. The protograph of this LDPC-chunked code can be expressed as
	\renewcommand\arraystretch{0.7}	
	\begin{equation}
		{\bf B}^{(1)} = \begin{bmatrix}
			1 0 0 0 &0 1 0 0 &1 0 0 0\\
			0  1  0  0 & 0  0  0  0 & 0  1  0  0\\
			0  1  0  0 & 0  1  0  0 & 0  0  1  0
		\end{bmatrix},
	\end{equation}
	\begin{equation}
		{\bf B}^{(2)} = \begin{bmatrix}
			1 1 1 1 & 0 0 0 0& 0 0 0 0 \\
			0 0 0 0& 1 1 1 1 &  0 0 0 0\\
			0 0 0 0& 0 0 0 0&  1 1 1 1
		\end{bmatrix}.
	\end{equation}
	\renewcommand\arraystretch{1}	
\end{example}

\subsection{Overlapped Chunked Codes}
Overlapped chunked codes \cite{overlapChunk} can be regarded as LDPC-chunked codes that employs a \textit{repetition code} as the sparse precode. If all L-CNs in a P-BNC has degree two and the two coefficients in the check equation are fixed to the identity element in $\mathbb{F}$, this P-BNC has a repetition precode and becomes a protograph-based overlapped chunked code.

Let ${\bf B}^{(1)}$ be a protomatrix with row weight of 2. Note that the row weight of 2 merely imposes an overlapping constraint on VNs, but it does not specify the selection of the overlapped VNs. Heidarzadeh and Banihashemi \cite{overlapChunk} proposed to share a fixed fraction of packets between two consecutive chunks in an end-around fashion. Following their overlapping scheme, ${\bf B}^{(1)}$ can be expressed as a block circulant shift matrix. Define two $n_o \times L$ circulant shift matrices
\renewcommand\arraystretch{0.7}	
\begin{equation}
	{\bf R}_1 = \begin{bmatrix}
		{\bf e}_{L-n_o+1} \\
		{\bf e}_{L-n_o+2} \\
		\vdots \\
		{\bf e}_{L} \\
	\end{bmatrix},~~~{\bf R}_2 = \begin{bmatrix}
		{\bf e}_1\\
		{\bf e}_2\\
		\vdots \\
		{\bf e}_{n_o}\\
	\end{bmatrix},
\end{equation}
\renewcommand\arraystretch{1}	
where ${\bf e}_i$ is a length-$L$ all-zero row vector except that the $i$-th component is one. If every two consecutive chunks have $n_o$ overlapped VNs, ${\bf B}^{(1)}$ becomes an $n_v/L \times n_v/L$ block circulant shift matrix
\renewcommand\arraystretch{0.7}	
\begin{equation}\label{eq:overlapped_B1}
	{\bf B}^{(1)} = \begin{bmatrix}
		{\bf R}_1 & {\bf R}_2 & \bf O & \cdots & \bf O\\
		\bf O & {\bf R}_1 & {\bf R}_2 & \cdots & \bf O\\
		\vdots & \vdots & \vdots & \ddots & \vdots \\
		{\bf R}_2 & \bf O & \bf O & \cdots & {\bf R}_1
	\end{bmatrix}_{\frac{n_v}{L} \times \frac{n_v}{L}},
\end{equation}
\renewcommand\arraystretch{1}	
where $\bf O$ is an $n_o \times L$ all-zero matrix. ${\bf B}^{(2)}$ is still defined as~(\ref{eq:L_chunk_B2}). 

Due to the lifting operation, an overlapped chunked code constructed from the protograph given above is not identical to the code described in \cite{overlapChunk}. However, the crucial feature that each chunk has a fixed fraction of overlapped packets is preserved.

\begin{example}
	Consider the overlapped chunked code with $L = 4$ and $n_o = 2$. The protomatrix of this overlapped chunked code can be written as
	\renewcommand\arraystretch{0.7}	
	\begin{equation}
		{\bf B}^{(1)} = \begin{bmatrix}
			0 0 1 0 &    1 0 0 0    & 0 0 0 0\\
			0  0  0  1 & 0  1  0  0 & 0  0  0  0\\
			0  0  0  0 & 0  0  1  0 & 1  0  0  0\\
			0  0  0  0 & 0  0  0  1 & 0  1  0  0\\
			1  0  0  0 & 0  0  0  0 & 0  0  1  0\\
			0  1  0  0 & 0  0  0  0 & 0  0  0  1\\
		\end{bmatrix},
	\end{equation}
	\begin{equation}
		{\bf B}^{(2)} = \begin{bmatrix}
			1 1 1 1 & 0 0 0 0& 0 0 0 0 \\
			0 0 0 0& 1 1 1 1 &  0 0 0 0\\
			0 0 0 0& 0 0 0 0&  1 1 1 1
		\end{bmatrix}.
	\end{equation}
	\renewcommand\arraystretch{1}	
\end{example}

\subsection{Gamma Codes}
Gamma codes \cite{GammaCodes2012} are similar to LDPC-chunked codes except that a fixed-rate Raptor code is specifically chosen to be the sparse precode. Here, we ignore the dense precode in the Raptor code and focus only on the LT code, as P-BNCs can also concatenate a dense precode after lifting. A fixed-rate LT code is equivalent to a low-density generator-matrix (LDGM) code with the information part being punctured. Thus, ${\bf B}^{(1)}$ is a systematic matrix as follows:
\begin{equation}
	{\bf B}^{(1)} = \begin{bmatrix}
		b_{1,1}^{(1)} & \cdots & b_{1,n_v-n_c^{(1)}}^{(1)} & 1 & 0 & \cdots & 0\\
		b_{2,1}^{(1)} & \cdots & b_{2,n_v-n_c^{(1)}}^{(1)} & 0 & 1 & \cdots & 0\\
		\vdots & \ddots & \vdots & \vdots & \vdots&\ddots & \vdots\\
		b_{n_c^{(1)},1}^{(1)} & \cdots & b_{n_c^{(1)},n_v-n_c^{(1)}}^{(1)} & 0 & 0 &0 &1\\
	\end{bmatrix}.
\end{equation}
Only the VNs corresponding to the identity matrix in ${\bf B}^{(1)}$ will be divided into chunks, each of size $L$. Suppose $n_c^{(1)}$ is a multiple of $L$. ${\bf B}^{(2)}$ is a juxtaposition of an $n_c^{(2)} \times (n_v-n_c^{(1)})$ all-zero matrix and an $({n_c^{(1)}}/{L}) \times ({n_c^{(1)}}/{L})$ block diagonal matrix as defined by (\ref{eq:L_chunk_B2}), where $n_c^{(2)} = {n_c^{(1)}}/{L}$.

\bibliographystyle{IEEEtran}
\bibliography{reference}

\begin{thebibliography}{10}
\providecommand{\url}[1]{#1}
\csname url@samestyle\endcsname
\providecommand{\newblock}{\relax}
\providecommand{\bibinfo}[2]{#2}
\providecommand{\BIBentrySTDinterwordspacing}{\spaceskip=0pt\relax}
\providecommand{\BIBentryALTinterwordstretchfactor}{4}
\providecommand{\BIBentryALTinterwordspacing}{\spaceskip=\fontdimen2\font plus
\BIBentryALTinterwordstretchfactor\fontdimen3\font minus
  \fontdimen4\font\relax}
\providecommand{\BIBforeignlanguage}[2]{{%
\expandafter\ifx\csname l@#1\endcsname\relax
\typeout{** WARNING: IEEEtran.bst: No hyphenation pattern has been}%
\typeout{** loaded for the language `#1'. Using the pattern for}%
\typeout{** the default language instead.}%
\else
\language=\csname l@#1\endcsname
\fi
#2}}
\providecommand{\BIBdecl}{\relax}
\BIBdecl

\bibitem{maymounkov2006methods}
P.~Maymounkov, N.~J. Harvey, D.~S. Lun \emph{et~al.}, ``Methods for efficient
  network coding,'' in \emph{Proc. 44th Annual Allerton Conference on
  Communication, Control, and Computing}, 2006, pp. 482--491.

\bibitem{5191397}
D.~Silva, W.~Zeng, and F.~R. Kschischang, ``Sparse network coding with
  overlapping classes,'' in \emph{Proc. Workshop on Network Coding, Theory, and
  Applications (NetCod)}, Lausanne, Switzerland, Jun. 2009, pp. 74--79.

\bibitem{overlapChunk}
A.~Heidarzadeh and A.~H. Banihashemi, ``Overlapped chunked network coding,'' in
  \emph{Proc. IEEE Inf. Theory Workshop (ITW)}, Cairo, Egypt, Jan. 2010, pp.
  1--5.

\bibitem{5695118}
Y.~Li, E.~Soljanin, and P.~Spasojevic, ``Effects of the generation size and
  overlap on throughput and complexity in randomized linear network coding,''
  \emph{IEEE Trans. Inf. Theory}, vol.~57, no.~2, pp. 1111--1123, Feb. 2011.

\bibitem{GammaCodes2012}
K.~Mahdaviani, M.~Ardakani, H.~Bagheri, and C.~Tellambura, ``Gamma codes: A
  low-overhead linear-complexity network coding solution,'' in \emph{Proc. Int.
  Symp. Net. Coding (NetCod)}, Cambridge, MA, USA, Jun. 2012, pp. 125--130.

\bibitem{Tang2012EOC}
B.~Tang, S.~Yang, Y.~Yin, B.~Ye, and S.~Lu, ``Expander graph based overlapped
  chunked codes,'' in \emph{Proc. IEEE Int. Symp. Inf. Theory}, Cambridge, MA,
  USA, Jul. 2012, pp. 2451--2455.

\bibitem{BATS}
S.~Yang and R.~W. Yeung, ``Batched sparse codes,'' \emph{IEEE Trans. Inf.
  Theory}, vol.~60, no.~9, pp. 5322--5346, Sept. 2014.

\bibitem{Tang2018Lchunked}
B.~Tang and S.~Yang, ``An {LDPC} approach for chunked network codes,''
  \emph{IEEE/ACM Trans. Networking}, vol.~26, no.~1, pp. 605--617, Feb. 2018.

\bibitem{Yang2016tree}
S.~Yang and Q.~Zhou, ``Tree analysis of {BATS} codes,'' \emph{IEEE Commun.
  Lett.}, Jan. 2016.

\bibitem{FL_analysis_BATS}
S.~Yang, T.-C. Ng, and R.~W. Yeung, ``Finite-length analysis of {BATS} codes,''
  \emph{IEEE Trans. Inf. Theory}, vol.~64, no.~1, pp. 322--348, Jan. 2018.

\bibitem{Xu2017QUBATS}
X.~Xu, Y.~L. Guan, Y.~Zeng, and C.-C. Chui, ``Quasi-universal {BATS} code,''
  \emph{IEEE Trans. Veh. Technol.}, vol.~66, no.~4, pp. 3497--3501, Apr. 2017.

\bibitem{8013842}
X.~Xu, Y.~Zeng, Y.~L. Guan, and L.~Yuan, ``Expanding-window {BATS} code for
  scalable video multicasting over erasure networks,'' \emph{IEEE Trans.
  Multimedia}, vol.~20, no.~2, pp. 271--281, Feb. 2018.

\bibitem{8629008}
J.~Yang, Z.-P. Shi, C.-X. Wang, and J.-B. Ji, ``Design of optimized
  sliding-window {BATS} codes,'' \emph{IEEE Commun. Lett.}, vol.~23, no.~3, pp.
  410--413, Mar. 2019.

\bibitem{9594262}
S.~Jayasooriya, J.~Yuan, and Y.~Xie, ``An improved sliding window {BATS}
  code,'' in \emph{Int. Symp. Topics in Coding (ISTC)}, Montreal, QC, Canada,
  Aug.-Sept. 2021, pp. 1--5.

\bibitem{9664430}
J.~Yang, Z.~Shi, and J.~Ji, ``Design of improved expanding-window {BATS}
  codes,'' \emph{IEEE Trans. Veh. Technol.}, vol.~71, no.~3, pp. 2874--2886,
  Mar. 2022.

\bibitem{Zhu2023LDPCBATS}
W.~Zhang, M.~Zhu, M.~Jiang, and N.~Hu, ``Design and optimization of {LDPC}
  precoded finite-length {BATS} codes under {BP} decoding,'' \emph{IEEE Commun.
  Lett.}, vol.~27, no.~12, pp. 3151--3155, Dec. 2023.

\bibitem{qing2024dependence}
J.~Qing, X.~Cai, Y.~Fan, M.~Zhu, and R.~W. Yeung, ``Dependence analysis and
  structured construction for batched sparse code,'' \emph{IEEE Trans.
  Commun.}, 2024.

\bibitem{Gallager}
R.~G. Gallager, \emph{Low-Density Parity-Check Codes}.\hskip 1em plus 0.5em
  minus 0.4em\relax The MIT Press, Sept. 1963.

\bibitem{910577}
T.~Richardson and R.~Urbanke, ``The capacity of low-density parity-check codes
  under message-passing decoding,'' \emph{IEEE Trans. Inf. Theory}, vol.~47,
  no.~2, pp. 599--618, Feb. 2001.

\bibitem{modern_coding_theory}
------, \emph{Modern Coding Theory}.\hskip 1em plus 0.5em minus 0.4em\relax New
  York, NY: Cambridge University Press, 2008.

\bibitem{zhu2024TIT}
M.~Zhu, S.~Yang, M.~Jiang, and C.~Zhao, ``Performance bounds and
  degree-distribution optimization of finite-length {BATS} codes,'' \emph{IEEE
  Trans. Inf. Theory}, 2025, to be published.

\bibitem{JPL}
J.~Thorpe, ``Low-density parity-check ({LDPC}) codes constructed from
  protographs,'' \emph{JPL IPN Progress Report}, pp. 42--154, Aug. 2003.

\bibitem{1523619}
D.~Divsalar, S.~Dolinar, and C.~Jones, ``Low-rate {LDPC} codes with simple
  protograph structure,'' in \emph{Proc. IEEE Int. Symp. Inf. Theory},
  Adelaide, SA, Australia, Sept. 2005, pp. 1622--1626.

\bibitem{5174517}
D.~Divsalar, S.~Dolinar, C.~R. Jones, and K.~Andrews, ``Capacity-approaching
  protograph codes,'' \emph{IEEE J. Sel. Areas Commun.}, vol.~27, no.~6, pp.
  876--888, Aug. 2009.

\bibitem{6266764}
T.~V. Nguyen, A.~Nosratinia, and D.~Divsalar, ``The design of rate-compatible
  protograph {LDPC} codes,'' \emph{IEEE Trans. Commun.}, vol.~60, no.~10, pp.
  2841--2850, Oct. 2012.

\bibitem{7045568}
T.-Y. Chen, K.~Vakilinia, D.~Divsalar, and R.~D. Wesel, ``Protograph-based
  raptor-like {LDPC} codes,'' \emph{IEEE Trans. Commun.}, vol.~63, no.~5, pp.
  1522--1532, May 2015.

\bibitem{7112076}
Y.~Fang, G.~Bi, Y.~L. Guan, and F.~C.~M. Lau, ``A survey on protograph {LDPC}
  codes and their applications,'' \emph{IEEE Communications Surveys \&
  Tutorials}, vol.~17, no.~4, pp. 1989--2016, 2015.

\bibitem{10639052}
D.~Simegn, K.~Andreev, P.~Rybin, and A.~Frolov, ``On the design of {LDPC}-based
  error-reducing codes,'' in \emph{Proc. Int. Symp. Wirel. Commun. Syst.
  (ISWCS)}, Rio de Janeiro, Brazil, Jul. 2024, pp. 1--6.

\bibitem{1176612}
S.-Y. Li, R.~Yeung, and N.~Cai, ``Linear network coding,'' \emph{IEEE Trans.
  Inf. Theory}, vol.~49, no.~2, pp. 371--381, Feb. 2003.

\bibitem{LOC2010}
S.~Yang, J.~Meng, and E.-h. Yang, ``Coding for linear operator channels over
  finite fields,'' in \emph{Proc. IEEE Int. Symp. Inf. Theory}, Austin, TX,
  USA, Jun. 2010.

\bibitem{LT}
M.~Luby, ``{LT} codes,'' in \emph{Proc. 43rd Annu. IEEE Symp. Found. Comput.
  Sci.}, Vancouver, BC, Canada, Nov. 2002, pp. 271--280.

\bibitem{Raptor}
A.~Shokrollahi, ``Raptor codes,'' \emph{IEEE Trans. Inf. Theory}, vol.~52,
  no.~6, pp. 2551--2567, Jun. 2006.

\bibitem{1705002}
T.~Ho, M.~M\'{e}dard, R.~Koetter, D.~Karger, M.~Effros, J.~Shi, and B.~Leong,
  ``A random linear network coding approach to multicast,'' \emph{IEEE Trans.
  Inf. Theory}, vol.~52, no.~10, pp. 4413--4430, Oct. 2006.

\bibitem{7110514}
B.~Tang, S.~Yang, B.~Ye, S.~Guo, and S.~Lu, ``Near-optimal one-sided scheduling
  for coded segmented network coding,'' \emph{IEEE Trans. Comput.}, vol.~65,
  no.~3, pp. 929--939, Mar. 2016.

\bibitem{8606284}
Z.~Zhou, C.~Li, S.~Yang, and X.~Guang, ``Practical inner codes for {BATS} codes
  in multi-hop wireless networks,'' \emph{IEEE Trans. Veh. Technol.}, vol.~68,
  no.~3, pp. 2751--2762, Mar. 2019.

\bibitem{yang2022bats}
S.~Yang and R.~W. Yeung, \emph{BATS Codes: Theory and practice}.\hskip 1em plus
  0.5em minus 0.4em\relax Morgan \& Claypool, 2022.

\bibitem{Raptor_codes_foundations_and_trends}
\BIBentryALTinterwordspacing
A.~Shokrollahi and M.~Luby, ``Raptor codes,'' \emph{Foundations and Trends® in
  Communications and Information Theory}, vol.~6, no. 3–4, pp. 213--322,
  2011. [Online]. Available: \url{http://dx.doi.org/10.1561/0100000060}
\BIBentrySTDinterwordspacing

\bibitem{PEG}
X.-Y. Hu, E.~Eleftheriou, and D.~Arnold, ``Regular and irregular progressive
  edge-growth {Tanner} graphs,'' \emph{IEEE Trans. Inf. Theory}, vol.~51,
  no.~1, pp. 386--398, Jan. 2005.

\bibitem{4411526}
G.~Liva and M.~Chiani, ``Protograph {LDPC} codes design based on {EXIT}
  analysis,'' in \emph{Proc. IEEE Global Telecom. Conf.}, Washington, DC, USA,
  Nov. 2007, pp. 3250--3254.

\bibitem{10056787}
Z.~Xiao, L.~Li, J.~Xu, and J.~Sha, ``Construction of protograph {LDPC} codes
  based on the convolution neural network,'' \emph{China Communications},
  vol.~20, no.~5, pp. 84--92, May 2023.

\bibitem{choukroun2024factorgraphoptimizationerrorcorrecting}
\BIBentryALTinterwordspacing
Y.~Choukroun and L.~Wolf, ``Factor graph optimization of error-correcting codes
  for belief propagation decoding,'' 2024. [Online]. Available:
  \url{https://arxiv.org/abs/2406.12900}
\BIBentrySTDinterwordspacing

\bibitem{7339431}
F.~Steiner, G.~Böcherer, and G.~Liva, ``Protograph-based {LDPC} code design
  for shaped bit-metric decoding,'' \emph{IEEE J. Sel. Areas Commun.}, vol.~34,
  no.~2, pp. 397--407, Feb. 2016.

\bibitem{6620513}
A.~K. Pradhan, A.~Subramanian, and A.~Thangaraj, ``Deterministic constructions
  for large girth protograph {LDPC} codes,'' in \emph{IEEE Int. Symp. Inf.
  Theory}, 2013, pp. 1680--1684.

\bibitem{my_github}
\BIBentryALTinterwordspacing
M.~Zhu, ``Protograph-based batched network codes implementation,'' Aug. 2024.
  [Online]. Available: \url{https://github.com/zhu-mingyang/bnc}
\BIBentrySTDinterwordspacing

\bibitem{Glebov}
A.~Glebov, L.~Medova, P.~Rybin, and A.~Frolov, ``On {LDPC} code based massive
  random-access scheme for the gaussian multiple access channel,'' in
  \emph{Proc. Internet of Things, Smart Spaces, and Next Generation Networks
  and Systems}.\hskip 1em plus 0.5em minus 0.4em\relax Cham: Springer
  International Publishing, 2018, pp. 162--171.

\bibitem{9797778}
A.~K. Pradhan, V.~K. Amalladinne, A.~Vem, K.~R. Narayanan, and J.-F.
  Chamberland, ``Sparse {IDMA}: A joint graph-based coding scheme for unsourced
  random access,'' \emph{IEEE Trans. Commun.}, vol.~70, no.~11, pp. 7124--7133,
  Nov. 2022.

\bibitem{6134051}
T.-Y. Chen, D.~Divsalar, J.~Wang, and R.~D. Wesel, ``Protograph-based
  raptor-like {LDPC} codes for rate compatibility with short blocklengths,'' in
  \emph{Proc. IEEE Global Telecommun. Conf.}, Houston, TX, USA, Dec. 2011, pp.
  1--6.

\bibitem{8422961}
M.~Ebada, A.~Elkelesh, S.~Cammerer, and S.~ten Brink, ``Scattered {EXIT} charts
  for finite length {LDPC} code design,'' in \emph{IEEE Int. Conf. Commun.
  (ICC)}, 2018, pp. 1--7.

\bibitem{8626506}
S.~V.~S. Ranganathan, D.~Divsalar, and R.~D. Wesel, ``Quasi-cyclic
  protograph-based raptor-like {LDPC} codes for short block-lengths,''
  \emph{IEEE Trans. Inf. Theory}, vol.~65, no.~6, pp. 3758--3777, 2019.

\bibitem{8846017}
A.~Elkelesh, M.~Ebada, S.~Cammerer, L.~Schmalen, and S.~ten Brink,
  ``Decoder-in-the-loop: Genetic optimization-based {LDPC} code design,''
  \emph{IEEE Access}, vol.~7, pp. 141\,161--141\,170, 2019.

\end{thebibliography}
\end{document}